\documentstyle[11pt,psfig,amssymb] {article}
\input epsf.sty

\newcommand{\be}{\begin{equation}}
\newcommand{\ee}{\end{equation}}
\newcommand{\bd}{\begin{displaymath}}
\newcommand{\ed}{\end{displaymath}}
\newcommand{\ba}{\begin{array}}
\newcommand{\ea}{\end{array}}
\newcommand{\bt}{\begin{tabular}}
\newcommand{\et}{\end{tabular}}
\newcommand{\bc}{\begin{center}}
\newcommand{\ec}{\end{center}}
\newcommand{\bn}{\begin{enumerate}}
\newcommand{\en}{\end{enumerate}}
\newcommand{\bi}{\begin{itemize}}
\newcommand{\ei}{\end{itemize}}
\newcommand{\bqr}{\begin{eqnarray}}
\newcommand{\eqr}{\end{eqnarray}}
\newcommand{\bfig}{\begin{figure}[tbp]}
\newcommand{\efig}{\end{figure}}
\newcommand{\btab}{\begin{table}[ht]}
\newcommand{\etab}{\end{tabular}\ec\end{table}}
\newcommand{\bl}{\begin{large}}
\newcommand{\el}{\end{large}}

\newcommand{\nb}{\nonumber}

\newcommand{\Ps}{P_{\sigma}}

\newcommand{\nuc}[2]{\mbox{\relax\ifmmode{}^{#1}{\protect\text{#2}}\else${}^{#1}$#2\fi}}

\title{Density Dependence of Two-Body Interactions \\
 for Beyond Mean-Field Calculations.}
\author     {
             T. Duguet\footnote{Present address: Argonne National Laboratory, Physics Division, 9700 S. Cass Av., Argonne, IL, 60439, USA. E-mail~: duguet@theory.phy.anl.gov} and P. Bonche, \\
             {\em Service de Physique Th\'eorique, CEA Saclay,} \\
             {\em 91191 Gif sur Yvette Cedex, France} \\
            }

\begin{document}

\maketitle

\begin{abstract}

This paper deals with the theoretical foundation of effective two-body forces for the Generator Coordinate Method (GCM) and the projected mean-field method. The first aim of this paper is to reduce into various local-densities the in-medium content of a generalized $G$ matrix removing the hard core problem in this extended context. Then, we consider the possible renormalization of multi-body forces through a density-dependent two-body interaction in the context of configuration mixing calculations. A density dependence of the form $\rho^{\sigma}$, as used in Skyrme and Gogny forces, is successfully interpreted as doing so when the mixed density is used. Finally, we propose a simple extension of the Skyrme force dedicated to the calculation of matrix elements between non-orthogonal product states, which are needed to evaluate the correlated energy.



\end{abstract}

\section{Introduction}
\label{intro}

Mean-field approximations are a first step towards the description of the ground state and low-lying excited states in microscopic nuclear structure calculations. They take into account the physical content of the one-body density matrix only. Consequently, data not related to one-body operators may not be satisfactorily reproduced in this context. Mean-field approximations make use of a product wave function to approximate the eigenstates of the systems in the variational problem. It is a Slater determinant in the Hartree-Fock (HF) theory~\cite{ring1} or a quasi-particle (qp) vacuum in the Hartree-Fock-Bogolyubov (HFB) theory~\cite{bogo} which includes static pairing correlations.

In cases where the agreement with the data is not satisfactory, one has to go beyond this approximation to include important missing correlations. Another reason to perform more elaborate calculations is the necessary restoration of symmetries possibly broken by mean-field solutions. When the symmetry breaking is weak, including appropriate correlations allows for a significant improvement of the binding energy in addition to obtain a wave-function with good quantum numbers.

Some variational methods beyond the mean-field make use of $N$-body wave-functions of the form:

\begin{equation}
| \, \Psi_{k} \rangle \, = \, \sum_{\alpha} f^{k}_{\alpha} \, | \, \Phi^{\alpha}_{0} \rangle \, \, \, \, \, \, ,  \label{mixing}
\end{equation}
where $\{| \, \Phi^{\alpha}_{0} \rangle\}$ constitutes a set of product states. This discrete superposition is sometimes an approximation for a mixing originally written in terms of a continuous coordinate $\alpha$. This is the case in the Generator Coordinate Method (GCM) and for some symmetry restorations. Once such a trial wave-function is given, its mean energy:

\begin{eqnarray}
{\cal E}_{k}^{mix}  &\equiv&  \, \, \, \, \, \, \, \,  \, \, \frac{\langle \, \Psi_{k} \, | \, H \, | \, \Psi_{k} \rangle}{\langle \, \Psi_{k} \, | \, \Psi_{k} \rangle} \, \, \, \,  \, \, ,  \label{factorisee} \\
\nb \\
&=& \frac{\sum_{\alpha,\beta} \, f_{\beta}^{k \, \ast} f^{k}_{\alpha} \langle \, \Phi^{\beta}_{0} \, | \, H \, | \, \Phi^{\alpha}_{0}  \rangle}{\sum_{\alpha,\beta} \, f_{\beta}^{k \, \ast} f^{k}_{\alpha} \langle \, \Phi^{\beta}_{0} \, | \, \Phi^{\alpha}_{0}  \rangle} \, \, \, \, \, \, .  
\label{energy}
\end{eqnarray}
can be minimized with respect to variational parameters. The mean-field approximation  is recovered from Eq.~\ref{mixing}-\ref{energy} if only one coefficient is non zero; its self-consistent version corresponding to the minimization of the energy with respect to the individual wave-functions. Within this general framework, let us explicit some methods mixing low-energy configurations.

The GCM~\cite{hill,bonche1,bonche2} deals with large-amplitude collective motions in nuclei and allows the evaluation of ground-state correlations, low-lying collective spectra and transition matrix elements associated with different vibrational modes or pairing vibrations. Within this method, the mixed states, generally determined through constrained self-consistent calculations, are fixed non-orthogonal mean-field states $| \, \Phi_{0}^{\alpha}  \rangle$. The energy is varied with respect to $f^{k \, \ast}_{\beta}$.

The projected mean-field method~\cite{yocco,diet,egido1,ring2} is another kind of mixing. It is used to restore symmetries broken at the mean-field level such as particle number and translational or rotational invariance. The degeneracy of the symmetry breaking mean-field solutions $| \, \Phi_{0}^{\alpha}  \rangle$ is used to construct eigenstates of the infinitesimal generator of the associated symmetry group. For Abelian groups such a requirement fixes the coefficients of the mixing $f^{k}_{\alpha}$~\cite{ring1} and the Variation After Projection (VAP) is performed with respect to the individual wave functions from which the non-orthogonal product states $| \, \Phi_{0}^{\alpha}  \rangle$ are built. 

Finally, an improvement of the ground-state as well as of low-lying state properties can be achieved by going beyond the independent quasi-particles picture through a mixing of quasi-particle excitations. This corresponds to the treatment of small amplitude correlations. The set of product states is composed of zero, two, four etc... quasi-particles states $| \, \Phi_{i}^{\alpha}  \rangle$ built on top of the mean-field state $| \, \Phi_{0}^{\alpha}  \rangle$ approximating the ground-state wave-function. This gives a set of fixed, orthonormal product functions. Alternatively, each of these quasi-particle states can be calculated self-consistently through the minimization of the corresponding mean-field energy. In this case, the mixed quasi-particle states $| \, \Phi_{i}^{\alpha}  \rangle$ are fixed non-orthogonal product wave-functions with identical quantum numbers. The variation is performed with respect to $f^{k \, \ast}_{i}$.

The former cases have been distinguished not only for their different physical content but also because one has in practice to determine whether the mixed product functions are orthogonal or not and what the variational parameters are. Indeed, the evaluation of the energy in the correlated state $| \, \Psi_{k} \rangle$ requires the calculation of matrix elements of the form $\langle \, \Phi^{\beta}_{0} \, | \, H \, | \, \Phi^{\alpha}_{0}  \rangle$. The evaluation of two, three\ldots body operators is feasible between non-orthogonal product states thanks to the generalized Wick theorem~\cite{bal1}. For orthogonal states one has to express both of them with respect to a single vacuum of reference before using the standard Wick theorem~\cite{wick}. As for the variation, a double variational method with respect to $f^{k \, \ast}_{i}$ and the single-particle wave-functions defining the mixed states could be performed in principle; at least when the coefficients of the mixing are not completely determined by geometrical arguments. However, the resolution of this problem is rather complicated and often contains a high degree of redundancy~\cite{ring1}. Finally, the above cases can be mixed through a more general ansatz for the correlated function $| \, \Psi_{k} \rangle$ in order to study the coupling between different physical effects (coupling of collective and single-particle degrees of freedom~\cite{didong,taj}, GCM in symmetry restored collective spaces~\cite{val1,rodrig1,dugnew}).

However, a formal problem arises when considering the calculation of the correlated energy in connection with effective density-dependent Hamiltonians $H\left[\rho\right]$. This is the case for nuclear structure calculations with phenomenological interactions~\cite{vauth,dech}. 

At the mean-field level, no ambiguity exists since the evaluation of Eq.~\ref{energy} requires the calculation of a single diagonal matrix element $\langle \, \Phi^{\alpha}_{0} \, | \, H\left[\rho\right] \, | \, \Phi^{\alpha}_{0}  \rangle$ and the one-body density used is naturally taken as the corresponding mean-field density. Precisely, it is generally taken as the local scalar-isoscalar part of the one-body density matrix~\cite{vauth,dech,chab}:

\begin{eqnarray}
\rho^{\alpha}_{0} (\vec{R}) \, \, &=& \, \, \frac{\langle \, \Phi^{\alpha}_{0} \, | \, \hat{\rho}_{0} (\vec{R}) \,  | \, \Phi^{\alpha}_{0}  \rangle}{\langle \, \Phi^{\alpha}_{0} \, | \, \Phi^{\alpha}_{0}  \rangle}  \label{usualdensity}  \, \, \, \, ,
\end{eqnarray}
where the local scalar-isoscalar part of the one-body density operator is defined as:

\begin{equation}
\hat{\rho}_{0} (\vec{r}) \, = \, \sum_{I,J,\zeta_{z},\zeta_{z}',s_{z},t} \, \varphi_{I}^{\ast} (\vec{r}, \zeta_{z}', s_{z},t) \, \varphi_{J} (\vec{r}, \zeta_{z}, s_{z}, t) \, \, c^{\dagger}_{I\zeta_{z}'s_{z}t} \, c_{J\zeta_{z}s_{z}t} \, \, \, \, .
\label{densityoperator}
\end{equation}

The individual wave-functions $\varphi_{I=N\pi} (\vec{r}, \zeta_{z}, s_{z}, t)$ have good parity $\pi$, $z$-signature $\zeta_{z}$, spin projection on the $z$-axis $s_{z}$ and isospin projection $t$; $N$ being the principal quantum number\footnote{Upper cases will denote throughout the paper all quantum numbers {\it except} for signature, spin and isospin while lower cases will denote all quantum numbers {\it including} signature, spin and isospin.}. They constitute a basis of the single-particle Hilbert space.

When going beyond the mean-field approximation through the use of the trial state $| \, \Psi_{k} \rangle$, there is no natural choice for the local density to insert into non-diagonal matrix elements $\langle \, \Phi^{\beta}_{0} \, | \, H\left[\rho\right] \, | \, \Phi^{\alpha}_{0}  \rangle$. Two prescriptions have been used up to now in such calculations:

\bi
\item[(1)] the local scalar-isoscalar part of the {\it mixed} density:
\begin{equation}
\rho^{\left(\beta,\alpha\right)}_{0} (\vec{R}) \, \, = \, \, \frac{\langle \, \Phi_{0}^{\beta} \, | \, \hat{\rho}_{0} (\vec{R}) \,  | \, \Phi_{0}^{\alpha}  \rangle}{\langle \, \Phi_{0}^{\beta} \, | \, \Phi_{0}^{\alpha}  \rangle} \, \, \, \, \, \, \, \, ,
\label{mixed}
\end{equation}
which has to be used in the corresponding kernel $\langle \, \Phi_{0}^{\beta} \, | \, H \, | \, \Phi_{0}^{\alpha}  \rangle$ in Eq.~\ref{energy}. This choice has been made in the GCM with or without projection on particle number and angular momentum~\cite{val1,rodrig1,dugnew,bonche3,haider,heenen}. Note that the diagonal mixed density is nothing but a  mean-field density~: $\rho^{\left(\alpha,\alpha\right)}_{0} (\vec{R}) = \rho^{\alpha}_{0} (\vec{R})$. So far, this choice has been motivated by the equivalence existing at the mean-field level between a three-body zero-range force and a linearly density-dependent two-body interaction~\cite{vauth} for time-reversal invariant systems. Considering this argument and the fact that the kernel $\langle \, \Phi_{0}^{\beta} \, | \, V^{(3)} \, | \, \Phi_{0}^{\alpha}  \rangle$ for a three-body force can be expressed in terms of the mixed density-matrix only, it was chosen to use the mixed local density in the two-body force~\cite{bonche3,haider,val2}. However, to our knowledge the above mentioned equivalence has not been shown explicitly for a general configuration mixing. The extension to a non-linear density-dependence as used in Skyrme and Gogny interactions has not been theoretically justified as well. 
\item[(2)] the local scalar-isoscalar part of the {\it correlated} density: 
\begin{equation}
\rho^{\Psi_{k}}_{0} (\vec{R}) \, \, = \, \, \frac{\langle \, \Psi_{k} \, | \, \hat{\rho}_{0} (\vec{R}) \, | \, \Psi_{k} \rangle}{\langle \, \Psi_{k} \, | \, \Psi_{k} \rangle} \, \, \, \, \, \, \, \, ,
\label{correlated}
\end{equation}
which does not depend on $(\beta,\alpha)$ and is to be used in all kernels in Eq.~\ref{energy}. This choice has been made in approximate and exact variation after particle number projection calculations~\cite{val2,val3,anguiano1}. Similar results to those found with the mixed density were obtained. The choice of the correlated density in the evaluation of the correlated energy seems as reasonable as the use of the mean-field density in the calculation of the mean-field energy. We discuss an a priori stronger argument in favor of the correlated density~\cite{val2} in appendix~\ref{argument}.
\ei

One could add the possibility to use the local mean-field density as defined by Eq.~\ref{usualdensity} in projected mean-field calculations. Indeed, it is possible in this case to express the correlated energy in terms of the density matrix and the pairing tensor of the single product state $| \, \Phi_{0}^{0}  \rangle$ from which $| \, \Psi_{k} \rangle$ is projected out~\cite{ring2}. This suggests the use of the corresponding mean-field density in the Hamiltonian in order to keep the energy as a functional of these variables only. No quantitative calculation has been performed until now using this prescription. It is worth noting that this argument does not hold in the general case. Indeed, the energy as given by Eq.~\ref{energy} cannot a priori be written in terms of the density matrix and pairing tensor of a single mean-field.

Given the lack of theoretical support for the commonly used prescriptions, the aim of the present work is to develop theoretical arguments in favor of a particular density dependence within the general framework of configuration mixing calculations. In order to do that, one has to identify physical reasons for the effective two-body force to depend on some density. We now give a non exhaustive review of the possible origins. 

\section{Density-Dependence Origins of Two-Body Interactions.}
\label{background}

One can roughly identify three cases where effective two-body forces should have a density dependence~\cite{bethe,nege3,mut}:

\bi
\item[(1)] the effective interaction includes some specific correlations induced by the bare nucleon-nucleon (NN) interaction through explicit summation of diagrams in a perturbative expansion. This is the case for two-body correlations associated with the repulsive core and the tensor component of the NN interaction which are taken care of, in presence of the other nucleons, through Brueckner ladders summation~\cite{brueck,day}. In that sense, phenomenological forces have been characterized as Brueckner $G$ matrices in the Local Density Approximation (LDA)~\cite{nege0,reid} or after a Density Matrix Expansion (DME)~\cite{nege}.

There are other correlations not considered explicitly at first-order in the $G$ matrix such as three-body, four-body... short-range correlations. The corresponding contributions to the binding energy can be phenomenologically included through a density dependence of the effective force~\cite{bethe}.

Long-range correlations can be treated explicitly through configuration mixings such as those considered in this work. If this is the case, the effective interaction must not renormalize them. 
\item[(2)] the phenomenological effective interaction omits explicitly some operators which are known to be important in realistic NN interactions and known to bring about a non trivial density-dependence in the binding energy. This is the case for the tensor force which is generally omitted in phenomenological interactions such as the Skyrme~\cite{vauth,chab} and the Gogny~\cite{dech} forces. As this component of the force plays a crucial role in the saturation process, any effective two-body interaction not including it explicitly should have a corresponding phenomenological density dependence~\cite{bethe2}. 
\item[(3)] the Hamiltonian includes the bare two-body force but higher-order interactions such as three-body terms are omitted. In this case, one can think of renormalizing their effects through a density-dependent two-body interaction~\cite{vauth,nege3}.
It is beyond the scope of this paper to discuss the sub-nucleonic origins of three-body and higher multi-body forces as well as their quantitative relevance~\cite{bethe,nege3}. However, it is worth mentioning their importance in nuclear systems. Three-body forces, either phenomenological or derived from microscopic meson-exchange models with nucleonic virtual excitations and nucleon-antinucleon pair creations, have been necessary to get good properties of (a)symmetric nuclear matter. In particular, they allow the reproduction of the empirical saturation at the correct density and energy per particle in non-relativistic Brueckner-Hartree-Fock calculations~\cite{grange,zuo}. Three-body forces have also been essential to describe spectroscopic properties of light nuclei up to mass $10$ in ab-initio calculations using the Green's function Monte Carlo method~\cite{wiringa}. In addition to these studies on bound nuclear systems, including three-body forces in the treatment of nucleon induced deuteron break-up or nucleon-deuteron elastic scattering has resolved several inconsistencies with the data observed when using two-body forces only~\cite{kuros}.
\ei

The renormalization of the described effects is understood at the mean-field level. The deduced effective two-body interaction always depends on the density of the underlying Slater determinant. However, as soon as the model or the variational space is modified to include correlations explicitly, not only must the analytical structure of the density dependence be changed but the kind of density itself is no longer obvious. 

The derivation of this analytical structure associated to each of the above mentioned origins constitutes a difficult task at any level of approximation. This is beyond the scope of this work. We rather concentrate on the determination of the {\it type} of density to be used in relation to origins (1) and (3) in calculations beyond the mean-field. The corresponding analytical functional will have a more phenomenological character. We are looking for guidelines to the definition of phenomenological forces while the phenomenology embodied by their fit on a few nuclear data should allow us to smooth out the imperfections of the analytical derivation.

Section~\ref{gmat} recalls some results needed to extend the usual Goldstone-Brueckner perturbation theory to configuration mixing calculations~\cite{duguet}. From section~\ref{lda} to section~\ref{firstlevel1}, we deal with the local density approximation of the generalized Brueckner matrix emerging in that extended perturbation theory. These calculations concern the particle-hole channel of the interaction since the Brueckner ladders should not be summed in the pairing channel~\cite{migdal,nozieres,gorkov,henley,dickhoff}. Section~\ref{gcm} then deals with the renormalization of three-body forces within the GCM while section~\ref{proj} is devoted to the same problem for the projected mean-field method. We generalize our results to higher multi-body forces in section~\ref{generalisation} and make some comments in section~\ref{quasi-particleconfig} about the configuration mixing of individual excitations. We supplement this part of the work with appendices providing details of our calculations, remarks concerning the use of the correlated density and a discussion on the crucial role of rearrangement terms in the equations of motion. The corresponding calculations are presented without taking care of static pairing correlations but the results can be extended to the pairing channel as well. Applying our results to the standard Skyrme force, we propose in section~\ref{appphenforces2} an extension of its density dependence to be used in projected mean-field and generator coordinate methods. As an application, the generalized Skyrme force is shown to be well defined for symmetry restoration. Conclusions are given in section~\ref{Conclu}.

\section{Two-Body Correlations for Mixed Non-Orthogonal Vacua.}
\label{brueckner}

\subsection{Generalized Brueckner matrix.}
\label{gmat}

In Ref.~\cite{duguet}, a generalized perturbation theory written in terms of non-orthogonal Slater determinants  has been developed. Contrary to usual perturbation theories~\cite{gold,hug,bloch1,bloch2}, this method aims at writing the actual ground state of the system as a superposition of solutions originating from several non-orthogonal product states close in energy at $t = -\infty$. One obtains a perturbative series taking care of long-range correlations which cannot be included in a simple way by using an expansion referring to a single vacuum. The new expansion still faces the problem related to the hard core of the two-body nucleon-nucleon interaction $V$. Ref.~\cite{duguet} shows how to sum generalized particle-particle ladders through a newly defined Brueckner matrix removing the hard-core problem in this extended context. This effective interaction $G^{\left(\beta,\alpha\right)}$ is given in terms of two standard Brueckner matrices $G^{\beta}$ and $G^{\alpha}$ referring to two different vacua $| \, \Phi^{\beta}_{0} \rangle$ and $| \, \Phi^{\alpha}_{0} \rangle$:

\begin{equation}
G^{\left(\beta,\alpha\right)} (W_{\beta},W_{\alpha}) \, = \, G^{\beta} (W_{\beta}) \, V^{-1} \, G^{\alpha} (W_{\alpha}) \label{supergmatrix} \, \, \, \, \, \, \, \, ,
\end{equation}
where the standard $G^{\alpha}$ Brueckner matrix satisfies a self-consistent equation of the form~\cite{brueck,gold}:

\begin{eqnarray}
G^{\alpha} (W_{\alpha}) \, &=& \, V  \, \, \, + \, \, \, V \, \, \, \frac{Q^{\alpha}}{W_{\alpha} \, \, - \, \, h^{\alpha}_{0}} \, \, G^{\alpha} (W_{\alpha})  \, \, \, \,  \, \,  .  \label{gmatrix} 
\end{eqnarray}

$Q^{\alpha}$ is the Pauli operator acting in the two-particle space to exclude occupied states in $| \, \Phi^{\alpha}_{0} \rangle \,$ as intermediate states in the Brueckner ladder. Eqs.~\ref{supergmatrix} and~\ref{gmatrix} make use of the one-body Hamiltonian $h^{\alpha}_{0}$ whose $N$-body ground-state is $|  \,  \Phi^{\alpha}_{0}  \rangle$ (with energy ${\cal E}^{\alpha}_{0}$). The single-particle eigenstates and eigenenergies of $h^{\alpha}_{0}$ are defined as $\{\phi_{\alpha_{k}},\epsilon_{\alpha_{k}}\}$ where $k$ denotes the quantum numbers $\{N\pi\zeta_{z}t\}$. This set of quantum numbers covers the cases of triaxially deformed and/or time-reversal symmetry breaking mean-field solutions. The starting energies $W_{\alpha}$ and $W_{\beta}$ characterize the dependence of the in-medium interaction of two particles on the energy of the others. The diagrammatic content of $G^{\left(\beta,\alpha\right)}$, its matrix elements and the precise definitions of $W_{\alpha}$ and $W_{\beta}$ are given in Ref.~\cite{duguet}. 

Considering the lowest-order of the extended perturbation theory, the ground-state energy is approximated by~\cite{duguet}:

\begin{equation}
{\cal E}_{0}^{\, n=0} \, = \, \frac{\sum_{\alpha,\beta} \, f_{\beta}^{0 \, \ast}  f^{0}_{\alpha}   \langle \, \Phi^{\beta}_{0} | \, t \, + \, G^{\left(\beta,\alpha\right)} (0,0) \, | \,  \Phi^{\alpha}_{0}  \rangle}{\sum_{\alpha,\beta}^{} \, f_{\beta}^{0 \, \ast}  f^{0}_{\alpha}  \langle \, \Phi^{\beta}_{0} |  \,  \Phi^{\alpha}_{0}  \rangle} \, \, \, \, \, . 
\label{lastenergy}
\end{equation}

This is precisely the energy ${\cal E}_{0}^{mix}$ for a state mixing non-orthogonal vacua as used in the GCM or the projected mean-field method where the starting two-body interaction $V$ has been replaced by the regularized effective vertex on the energy shell ($W_{\alpha} = W_{\beta} = 0$).

\subsection{LDA for Standard Brueckner Matrices: definitions and goal.}
\label{lda}

The interaction $G^{\alpha}$ includes many-body effects through the operator $Q^{\alpha} / (W_{\alpha} - h^{\alpha}_{0})$. The value of $W_{\alpha}$ depends upon the location of the $G^{\alpha}$ interaction in a given graph. At that location, the matrix element $\langle \, \alpha_m \, \alpha_n | \,  G^{\alpha} \, | \, \alpha_p \, \alpha_q  \rangle$ depends upon $\epsilon_{\alpha_p} + \epsilon_{\alpha_q} - W_{\alpha}$ through the energy denominator of $Q^{\alpha} / (W_{\alpha} - h^{\alpha}_{0})$. The question adressed now is whether these in-medium effects can be reduced to a simpler dependence on various local densities.

The $G^{\alpha}$ matrix elements are well defined in the $\{\phi_{\alpha_{k}}\}$ representation. Considering the transformation to configuration space, one can write:

\begin{eqnarray}
\langle \, \alpha_m \, \alpha_n \, | \,  G^{\alpha} \, | \, \alpha_p \, \alpha_q  \rangle &=& \sum_{\zeta_{z 1}\ldots t_{1}\ldots} \int \, d\vec{r}_{1} \, d\vec{r}_{2} \, d\vec{r'}_{1} \, d\vec{r'}_{2} \, \langle \, \alpha_m \, \alpha_n \, | \, \vec{r}_{1} \, \zeta_{z 1} \, t_{1} \,  \vec{r}_{2} \, \zeta_{z 2} \, t_{2}  \rangle \, \nb  \\
&& \label{matrixelement1} \\
&&  \langle \, \vec{r}_{1} \, \zeta_{z 1} \, t_{1} \,  \vec{r}_{2} \, \zeta_{z 2} \, t_{2} \, | \, G^{\alpha} \, | \, \vec{r'}_{1} \, \zeta'_{z 1} \, t'_{1} \,  \vec{r'}_{2} \, \zeta'_{z 2} \, t'_{2}  \rangle \, \langle \, \vec{r'}_{1} \, \zeta'_{z 1} \, t'_{1} \,  \vec{r'}_{2} \, \zeta'_{z 2} \, t'_{2} \,  | \, \alpha_p \, \alpha_q  \rangle \, \, \, \, \, , \nb
\end{eqnarray}
where $\langle \, \vec{r'}_{1} \, \zeta'_{z 1} \, t'_{1} \,  \vec{r'}_{2} \, \zeta'_{z 2} \, t'_{2} \,  | \, \alpha_p \, \alpha_q  \rangle$ is an antisymmetrized two-body wave function. Eq.~\ref{gmatrix} satisfied by the standard $G^{\alpha}$ matrix reads in configuration space as:

\begin{eqnarray}
&& \langle \, \vec{r}_{1} \, \zeta_{z 1} \, t_{1} \,  \vec{r}_{2} \, \zeta_{z 2} \, t_{2} \, | \, G^{\alpha} \, | \, \vec{r'}_{1} \, \zeta'_{z 1} \, t'_{1} \,  \vec{r'}_{2} \, \zeta'_{z 2} \, t'_{2}  \rangle \, = \, \langle \, \vec{r}_{1} \, \zeta_{z 1} \, t_{1} \,  \vec{r}_{2} \, \zeta_{z 2} \, t_{2} \, | \, V \, | \, \vec{r'}_{1} \, \zeta'_{z 1} \, t'_{1} \,  \vec{r'}_{2} \, \zeta'_{z 2} \, t'_{2}  \rangle \,  \nb \\
&& \label{matrixelement2} \\
&& \hspace{5cm} + \, \langle \, \vec{r}_{1} \, \zeta_{z 1} \, t_{1} \,  \vec{r}_{2} \, \zeta_{z 2} \, t_{2} \, | \, V \frac{Q^{\alpha}}{W_{\alpha} - h^{\alpha}_{0}} \, G^{\alpha} \, | \, \vec{r'}_{1} \, \zeta'_{z 1} \, t'_{1} \,  \vec{r'}_{2} \, \zeta'_{z 2} \, t'_{2}  \rangle \, , \nb 
\end{eqnarray}
and is linked to the original matrix element in the $\{\phi_{\alpha_{k}}\}$ representation through its dependence on $\epsilon_{\alpha_{p}} + \epsilon_{\alpha_{q}}$. Note that it is not necessary to antisymmetrize the matrix elements in coordinate space since it is done in the $\{\phi_{\alpha_{k}}\}$ representation. We now write the bare interaction under the form~\cite{ring1}:

\begin{equation}
\langle \vec{r} \, '_{1} \, \vec{r} \, '_{2}  | \, V \, | \vec{r}_{1} \, \vec{r}_{2} \,  \rangle \, = \, V (\vec{r}, \vec{p}, \hat{\vec{\sigma}}_{1}, \hat{\vec{\sigma}}_{2}, \hat{\vec{\tau}}_{1}.\hat{\vec{\tau}}_{2}) \, \, \delta (\vec{r}_{1} - \vec{r} \, '_{1}) \, \, \delta (\vec{r}_{2} - \vec{r} \, '_{2}) \, \, \, ,
\label{forcenue}
\end{equation}
where $\vec{r} = \vec{r}_{1} - \vec{r}_{2}$ and $\vec{p} = (\vec{p}_{1} - \vec{p}_{2})/2$ are the relative position and momentum vectors of the two interacting nucleons. The $\vec{p}$ dependence is a practical manner to take the non-locality of the force into account. Using this expression and the fact that $Q^{\alpha} / (W_{\alpha} - h^{\alpha}_{0})$ is diagonal in the (non-antisymmetrized here) two-particle basis $\left\{| \, \alpha_r \, \alpha_s  \rangle\right\}$~:

\begin{equation}
\frac{Q^{\alpha}}{W_{\alpha} - h^{\alpha}_{0}} \, =  \, \sum_{\mathrel{\mathop{\epsilon_{\alpha_{r}}  > \epsilon^{\alpha}_{F}}\limits_{\epsilon_{\alpha_{s}}  > \epsilon^{\alpha}_{F}}}}^{} \, \frac{|\, \alpha_{r} \alpha_{s}  \rangle \,  \langle \, \alpha_{r} \alpha_{s} \, |}{\epsilon_{\alpha_{p}} + \epsilon_{\alpha_{q}} - W_{\alpha} - \epsilon_{\alpha_{r}} - \epsilon_{\alpha_{s}}} \, \, \, ,
\label{projecteur1}
\end{equation}
one obtains by inserting the identity operator twice through closure relations in the last term of Eq~\ref{matrixelement2}:

\begin{eqnarray}
&&\langle \, \vec{r}_{1} \, \zeta_{z 1} \, t_{1} \,  \vec{r}_{2} \, \zeta_{z 2} \, t_{2} \, | \, V \frac{Q^{\alpha}}{W_{\alpha} - h^{\alpha}_{0}} \, G^{\alpha} \, | \, \vec{r'}_{1} \, \zeta'_{z 1} \, t'_{1} \,  \vec{r'}_{2} \, \zeta'_{z 2} \, t'_{2}  \rangle =  \nb \\
&& \nb \\
&&\sum_{\zeta_{z 3}\zeta_{z 4}t_{3}t_{4}} \langle \, \zeta_{z 1} \, t_{1} \,  \zeta_{z 2} \, t_{2} \, | \, V (\vec{r}, \vec{p}, \hat{\vec{\sigma_{1}}}, \hat{\vec{\sigma_{2}}}, \hat{\vec{\tau_{1}}}.\hat{\vec{\tau_{2}}}) \, | \, \zeta_{z 3} \, t_{3} \, \zeta_{z 4} \, t_{4}  \rangle \, \int \, d\vec{r}_{3} \, d\vec{r}_{4} \label{matrixelement3} \\
&& \nb \\
&& \langle \, \vec{r}_{1} \, \zeta_{z 3} \, t_{3} \,  \vec{r}_{2} \, \zeta_{z 4} \, t_{4} \,  | \, \frac{Q^{\alpha}}{W_{\alpha} - h^{\alpha}_{0}} \, | \,  \vec{r}_{3} \, \zeta_{z 3} \, t_{3} \,  \vec{r}_{4} \, \zeta_{z 4} \, t_{4}  \rangle \, \langle \,  \vec{r}_{3} \, \zeta_{z 3} \, t_{3} \,  \vec{r}_{4} \, \zeta_{z 4} \, t_{4} \,  | \,  G^{\alpha}  \, | \, \vec{r'}_{1} \, \zeta'_{z 1} \, t'_{1} \,  \vec{r'}_{2} \, \zeta'_{z 2} \, t'_{2}  \rangle  \, \, \, . \nb
\end{eqnarray}

\subsection{LDA for Standard Brueckner Matrices: analysis.}
\label{Analysis}

We study the operator $Q^{\alpha} / \, (W_{\alpha} - h^{\alpha}_{0})$ which is responsible for the in-medium effects and thus for the possible density dependence. Its non-antisymmetrized matrix elements read in coordinate space as:

\begin{equation}
\langle \, \vec{r}_{1} \, \zeta_{z 3} \, t_{3} \,  \vec{r}_{2} \, \zeta_{z 4} \, t_{4} \,  | \, \frac{Q^{\alpha}}{W_{\alpha} - h^{\alpha}_{0}} \, | \,  \vec{r}_{3} \, \zeta_{z 3} \, t_{3} \,  \vec{r}_{4} \, \zeta_{z 4} \, t_{4}  \rangle \, =   \, \sum_{\mathrel{\mathop{\epsilon_{\alpha_{r}}  > \epsilon^{\alpha}_{F_{t_{3}\zeta_{z 3}}}}\limits_{\epsilon_{\alpha_{s}}  > \epsilon^{\alpha}_{F_{t_{4}\zeta_{z 4}}}}}}^{}  \, \frac{ \phi^{\ast}_{\alpha_s} (\vec{r}_{4}) \, \phi_{\alpha_s} (\vec{r}_{2}) \,  \phi^{\ast}_{\alpha_r} (\vec{r}_{3}) \, \phi_{\alpha_r} (\vec{r}_{1}) \,}{\epsilon_{\alpha_{p}} + \epsilon_{\alpha_{q}} - W_{\alpha} - \epsilon_{\alpha_{r}} - \epsilon_{\alpha_{s}}} \, \, \, ,
\label{projecteur2}
\end{equation}
and are diagonal in isospin. The isospins of the states involved in the sum satisfy $(t_{r}=t_{3}, t_{s}=t_{4})$ and  $(\zeta_{z r}=\zeta_{z 3}, \zeta_{z s}=\zeta_{z 4})$. This is specified through two labelled Fermi energies $\epsilon^{\alpha}_{F_{t_{3}\zeta_{z 3}}}$ and $\epsilon^{\alpha}_{F_{t_{4}\zeta_{z 4}}}$.

In order to write an approximation in terms of local densities, one has to make the one-body density-matrix appear in the numerator of Eq.~\ref{projecteur2}. To do so, one has to perform some average on the energy denominator to make it independent of the running indices $(\alpha_{r}, \alpha_{s})$. In most existing works the procedure consists of averaging this denominator over the Fermi sea in nuclear matter~\cite{nege,nege2}. This mean value depends on the Fermi energy, which in turn depends upon the total density of the system. This dependence is eventually transformed into a local-density dependence when going back to finite nuclei. Together with a density-matrix expansion in the numerator, this provides the full local density dependence of the effective interaction~\cite{nege,nege2}. However, this strategy is doubtful from a formal point of view: indeed, three types of energies should be distinguished in the energy denominator of Eq.~\ref{projecteur2}.

The first energy, $\epsilon_{\alpha_{p}} + \epsilon_{\alpha_{q}}$, refers to the unperturbed two-body ket in the original matrix element~\ref{matrixelement1}. This energy is fixed in the right hand-side of Eq.~\ref{projecteur2}. It is indeed reasonable to average it out over the Fermi sea since the matrix elements of $G^{\alpha}$ involved in the calculation of the energy at the lowest order concern occupied states. However, such an average is related to the Fermi energies in the studied finite nucleus, and thus to $N$ and $Z$ rather than to a local density at the center of mass of the interacting nucleons. Besides, even if he used it successfully, Negele has shown the crudeness of the above procedure for this energy dependence~\cite{nege2}.

The second energy, $W_{\alpha}$, induces a non-trivial dependence of $G^{\alpha}$ on its location in the diagram. Given that location, $W_{\alpha}$ is fixed for all matrix elements defined by Eq.~\ref{matrixelement1}. At the lowest order, this parameter is zero since the Brueckner matrix is taken on the energy shell.

The last energy, $\epsilon_{\alpha_{r}} + \epsilon_{\alpha_{s}}$, makes the denominator of Eq.~\ref{projecteur2} dependent on the running indices $\alpha_{r}$ and $\alpha_{s}$. This energy comes from each two-body propagator in the Brueckner ladder and the indices run over particles states up to infinite energies. It is in no way related to any energy in the Fermi sea. As it is a weighting factor for each particle-particle intermediate state entering the ladder, it is strongly connected to the numerator in Eq.~\ref{projecteur2}. Omitting this link amounts to giving up an important part of the physics involved in the Brueckner resummation by providing particle states with the same weight whatever their unperturbed energies. 

The usual averaging procedure finally amounts to stating that the Pauli operator:

\begin{equation}
\langle \, \vec{r}_{1} \, \zeta_{z 3} \, t_{3} \,  \vec{r}_{2} \, \zeta_{z 4} \, t_{4} \,  | \, Q^{\alpha} \, | \,  \vec{r}_{3} \, \zeta_{z 3} \, t_{3} \,  \vec{r}_{4} \, \zeta_{z 4} \, t_{4}  \rangle = \left[\delta (\vec{r}_{2} - \vec{r}_{4}) - \rho^{\alpha}_{t_{4}\zeta_{z 4}}(\vec{r}_{2}, \vec{r}_{4})\right] \left[\delta (\vec{r}_{1} - \vec{r}_{3}) - \rho^{\alpha}_{t_{3}\zeta_{z 3}}(\vec{r}_{1}, \vec{r}_{3})\right] \, \, \, \, ,
\label{projecteur3}
\end{equation}
is the essential quantity to treat in detail in the operator $Q^{\alpha} / \, (W_{\alpha} - h^{\alpha}_{0})$. It is where the DME is performed~\cite{nege,nege2}. As the procedure providing Eq.~\ref{projecteur3} is questionable, and as a simplification of the $G^{\alpha}$ is by far necessary to proceed to extensive nuclear structure calculations, we could think of a direct local approximation of the ``energy-density'' defined by Eq.~\ref{projecteur2}~\cite{these}. Following the spirit of the DME, this approximation should be of similar quality as for Eq.~\ref{projecteur3}~\cite{nege}. However, this would not provide the interaction with dependences on local densities $(\rho^{\alpha}(\vec{R}), \nabla^2 \, \rho^{\alpha}(\vec{R}), \tau^{\alpha}(\vec{R})\ldots)$\footnote{For the definitions of the different densities, see for instance Ref.~\cite{engel}.}.

In spite of the weaknesses we have just discussed, there is no fully coherent picture available approximating the matrix element~\ref{projecteur2} by a functional of local densities. Consequently, we give up the idea of deriving analytically the density dependence induced by the operator $Q^{\alpha} / \, (W_{\alpha} - h^{\alpha}_{0})$.

\subsection{Ansatz.}
\label{Ansatz}

Because of the conclusions of the previous section, we {\it postulate} the possibility to approximate the matrix element~\ref{projecteur2} through a local, zero-range quantity of the form:

\begin{eqnarray}
\langle \, \vec{r}_{1} \, \zeta_{z 3} \, t_{3} \,  \vec{r}_{2} \, \zeta_{z 4} \, t_{4} \,  | \, \frac{Q^{\alpha}}{W_{\alpha} - h^{\alpha}_{0}} \, | \,  \vec{r}_{3} \, \zeta_{z 3} \, t_{3} \,  \vec{r}_{4} \, \zeta_{z 4} \, t_{4}  \rangle\!&\approx&\!{\cal K} \left[\rho^{\alpha}_{t_{3}\zeta_{z 3}}(\vec{r}_{1}), \rho^{\alpha}_{t_{4}\zeta_{z 4}}(\vec{r}_{2})\right] \, \delta (\vec{r}_{1} - \vec{r}_{3}) \, \delta (\vec{r}_{2} - \vec{r}_{4}) \nb \\
&& \label{projecteur5} \\
&\approx&\!{\cal F} \left[\rho_{0}^{\alpha} (\vec{R})\right] \, \delta (\vec{r}_{1} - \vec{r}_{3}) \, \delta (\vec{r}_{2} - \vec{r}_{4}) \, \delta (\vec{r}_{1} - \vec{r}_{2})  \, \, \, , \nb
\end{eqnarray}
where ${\cal F}$ (${\cal K}$) is an unknown functional of the local scalar-isoscalar (isospin and signature dependent) density(ies) associated with the product state $|   \Phi^{\alpha}_{0}  \rangle$. 

The DME, which we keep as a motivation for the local approximation~\ref{projecteur5}, suggests a dependence of ${\cal F}$ (${\cal K}$) on a power series of the relative coordinates $\vec{r}_{1} - \vec{r}_{3}$ and $\vec{r}_{2} - \vec{r}_{4}$~\cite{nege2} figuring the non-locality of the operator $Q^{\alpha} / \, (W_{\alpha} - h^{\alpha}_{0})$. Thus, some corrective terms depending on second derivatives of the local densities $\rho^{\alpha}(\vec{R})$ and on the kinetic densities $\tau^{\alpha}(\vec{R})$ should also be considered. For systems breaking time reversal invariance, the DME provides dependences on spin densities $\vec{s}^{ \, \alpha} (\vec{R})$, current densities $\vec{j}^{\alpha} (\vec{R})$ and spin-current densities $\vec{J}^{\alpha} (\vec{R})$. The fully local approximation embodied by Eq.~\ref{projecteur5} is correct at high density but particularly crude at the surface in finite nuclei. The zero-range approximation in $\vec{r}_{1} - \vec{r}_{2}$ is much safer since $Q^{\alpha} / \, (W_{\alpha} - h^{\alpha}_{0})$ is always surrounded by two $V$ interactions in $G^{\alpha}$~\cite{nege}. It is responsible for the dependence of $\rho_{t_3\zeta_{z 3}}^{\alpha}$ and $\rho_{t_4\zeta_{z 4}}^{\alpha}$ on $\vec{R}$ rather than on $\vec{r}_{1}$ and $\vec{r}_{2}$ respectively in Eq.~\ref{projecteur5}.

As suggested by Eqs.~\ref{projecteur3} and~\ref{projecteur5}, the in-medium content of the effective force depends on the isospin of the interacting nucleons and should provide a dependence on both the isoscalar and the isovector components of the local part of the density matrix. However, phenomenological forces used generally depend on the isoscalar density only whatever the isospin of the interacting nucleons. This has been satisfying for nuclei around the stability line but could be questionable for near drip-line nuclei. In the same spirit, the in-medium content of the effective force depends on the signature of the interacting nucleons and should provide a dependence on both the scalar and the vector components of the local part of the density matrix. This would differenciate the effective interaction for even-even and odd-even nuclei as well as for $J=0$ and $J>0$ states.

Taking the complete dependences into account would complicate the following theoretical and numerical developments dramatically. This is why none of the refinements with respect to a dependence of the local scalar-isoscalar part of the density matrix will be considered here. This is consistent with the present form of Gogny and Skyrme forces. Corrections will be included through the phenomenological nature of these interactions.

Using the ansatz~\ref{projecteur5}, and the fact that the operator $\vec{p} = (\vec{p}_{1}-\vec{p}_{2})/2$ does not act on $\vec{R}$, the matrix element defined through Eq.~\ref{matrixelement2} becomes~:

\begin{eqnarray}
&& \langle \, \zeta_{z 1} \, t_{1} \, \zeta_{z 2} \, t_{2} \, | \, G^{\alpha}_{LDA} (\vec{R}, \vec{r}, \vec{p}, \hat{\vec{\sigma_{1}}}, \hat{\vec{\sigma_{2}}}, \hat{\vec{\tau_{1}}}.\hat{\vec{\tau_{2}}}) - V (\vec{r}, \vec{p}, \hat{\vec{\sigma_{1}}}, \hat{\vec{\sigma_{2}}}, \hat{\vec{\tau_{1}}}.\hat{\vec{\tau_{2}}}) \, | \, \zeta'_{z 1} \, t'_{1} \, \zeta'_{z 2} \, t'_{2}  \rangle \, = \nb \\
&& \label{matrixelement5} \\
&&{\cal F} \left[\rho_{0}^{\alpha}(\vec{R})\right] \, \langle \, \zeta_{z 1} \, t_{1} \,  \zeta_{z 2} \, t_{2} \, | \, V (\vec{r}, \vec{p}, \hat{\vec{\sigma_{1}}}, \hat{\vec{\sigma_{2}}}, \hat{\vec{\tau_{1}}}.\hat{\vec{\tau_{2}}}) \, \, G^{\alpha}_{LDA} (\vec{R}, \vec{r}, \vec{p}, \hat{\vec{\sigma_{1}}}, \hat{\vec{\sigma_{2}}}, \hat{\vec{\tau_{1}}}.\hat{\vec{\tau_{2}}}) \, | \, \zeta'_{z 1} \, t'_{1} \, \zeta'_{z 2} \, t'_{2}  \rangle   \, \, \, . \nb
\end{eqnarray}

Let us introduce a closure relation in the tensorial product space of signature/isospin spaces between $V$ and $G^{\alpha}_{LDA}$, iterate Eq.~\ref{matrixelement5} and eliminate this closure relation. We obtain an expression valid for all $G^{\alpha}_{LDA}$ matrix elements in signature/isospin spaces which, without writting explicitly the dependences of $V$ and $G^{\alpha}_{LDA}$ on the operators $\hat{\vec{\sigma_{1}}}$, $\hat{\vec{\sigma_{2}}}$ and $\hat{\vec{\tau_{1}}}.\hat{\vec{\tau_{2}}}$, takes the form:

\begin{eqnarray}
G^{\alpha}_{LDA} (\vec{R}, \vec{r}, \vec{p}) \, &=& \,  \sum_{l=0}^{\infty} \, V^{l+1} (\vec{r}, \vec{p}) \, {\cal F}^{l} \left[\rho_{0}^{\alpha}(\vec{R})\right]  \nb \\
&& \label{ldagalpha} \\
&=& \, V (\vec{r}, \vec{p}) \, + \, V^{2} (\vec{r}, \vec{p}) \, {\cal F} \left[\rho_{0}^{\alpha}(\vec{R})\right] \,  + \, V^{3} (\vec{r}, \vec{p}) \, {\cal F}^{2} \left[\rho_{0}^{\alpha}(\vec{R})\right] \,  + \, \ldots  \nb 
\end{eqnarray}

\subsection{LDA for $G^{\left(\beta,\alpha\right)}$.}
\label{LDAGbetaalpha}

We now turn to the local approximation of the $G^{\left(\beta,\alpha\right)}$ matrix on the energy shell. Using Eq.~\ref{supergmatrix} and~Eq.~\ref{ldagalpha}, we obtain after some manipulations~:

\begin{eqnarray}
G_{LDA}^{\left(\beta,\alpha\right)} \, (\vec{R}, \vec{r}, \vec{p})  \, &=& \, \sum_{l=1}^{\infty} \, V^{l} (\vec{r}, \vec{p}) \, \frac{{\cal F}^{l} \left[\rho_{0}^{\beta}(\vec{R})\right] \, - \, {\cal F}^{l} \left[\rho_{0}^{\alpha}(\vec{R})\right]}{{\cal F} \left[\rho_{0}^{\beta}(\vec{R})\right] \, - \, {\cal F} \left[\rho_{0}^{\alpha}(\vec{R})\right]} \nb \\
&& \label{ldagbetaalpha} \\
&=& \, \, \, \, \, \, \, \, V (\vec{r}, \vec{p}) \, \, \, + \, \, \, V^{2} (\vec{r}, \vec{p}) \, \left\{{\cal F} \left[\rho_{0}^{\beta}(\vec{R})\right] \, + \, {\cal F} \left[\rho_{0}^{\alpha}(\vec{R})\right]\right\} \,   \nb \\
&& \nb  \\
&&\!+ \, V^{3} (\vec{r}, \vec{p}) \, \left\{{\cal F}^{2} \left[\rho_{0}^{\beta}(\vec{R})\right] \, + \, {\cal F} \left[\rho_{0}^{\beta}(\vec{R})\right] \, {\cal F} \left[\rho_{0}^{\alpha}(\vec{R})\right] \,  + \, {\cal F}^{2} \left[\rho_{0}^{\alpha}(\vec{R})\right]\right\} \, + \, \ldots  \nb
\end{eqnarray}

This form of the effective interaction relies on general manipulations only and is still far too complicated to be used in extensive calculations of finite nuclei. It needs to be simplified. However, one can already conclude one important point. Eq.~\ref{ldagbetaalpha} shows that whatever the explicit form of the functional ${\cal F}$ is, $G_{LDA}^{\left(\beta,\alpha\right)}$ will depend neither on the mixed density nor on the correlated density, but on local mean-field densities $\rho_{0}^{\alpha}(\vec{R})$ and $\rho_{0}^{\beta}(\vec{R})$ of the two product states involved in the matrix element $\langle \, \Phi_{0}^{\beta} | \, t + G^{\left(\beta,\alpha\right)} \, | \,\Phi_{0}^{\alpha}   \rangle$.

We shall now go through some simplifications. As standard phenomenological forces such as Gogny and Skyrme forces have often been interpreted as local density approximations of Brueckner matrices~\cite{vauth,nege2}, one should recover their usual mean-field density-dependence when considering a diagonal term $(\beta = \alpha)$ in Eq.~\ref{ldagbetaalpha}. For instance, a typical Skyrme force reads as~\cite{chab}:

\begin{eqnarray}
v_{Skyrme} (\vec{R}, \vec{r}, \overrightarrow{k}, \overleftarrow{k'}) &=& t_0 \, ( 1 +  x_0 \Ps ) \, \,  \delta (\vec{r})   \nb  \\
        && + \, \frac{1}{2} \, \, t_1 \, \,  ( 1  +  x_1  \Ps )  \,  \, (\delta (\vec{r}) \,  \, \overrightarrow{k}^2 \, + \, \overleftarrow{k'}^2 \,  \, \delta (\vec{r}) )   \nb \\
        && + \, \, t_2 \, \, ( 1  +  x_2  \Ps ) \,  \, \overleftarrow{k'} . \, \,  \delta (\vec{r}) \, \,  \overrightarrow{k} \label{skyrme}  \\
        && + \, \frac{1}{6} \, \, t_3 \, \, ( 1  +  x_3 \Ps ) \,  \, \left[\rho^{\alpha}_{0} (\vec{R})\right]^{\sigma} \, \, \delta (\vec{r})  \nb   \\
        && + \, \, i W_0  \, \, (\hat{\vec{\sigma_1}} \, + \, \hat{\vec{\sigma_2}} ) \, \, \overleftarrow{k'} \wedge \,  \, \delta (\vec{r}) \,  \, \overrightarrow{k}  \, \, \, \, , \nb
\end{eqnarray}
where $\overrightarrow{k} = (\nabla_{1} - \nabla_{2}) / 2i$ embodies the momentum operator acting on the right while $\overleftarrow{k'}$ embodies the same operator, with a sign minus, acting on the left. $\Ps$ is the spin exchange operator. Tensor forces are still not included in Skyrme forces. On one hand, Hartree-Fock calculations including such a term have shown no improvement on spin-orbit splittings in spin unsaturated nuclei~\cite{stancu}. On the other hand, perturbative calculations have shown the importance of the tensor component in generating two-body correlations and associated density-dependence~\cite{kohl1,kohl2}. This is due to the dependence of the tensor force contribution to the energy on the Pauli operator $Q^{\alpha}$. This remark is also supported by many-body calculations taking into account the mesonic degree of freedom which shows the strong in-medium effects generated by pions exchange~\cite{weise}. Thus, one would expect a density-dependent tensor term in phenomenological effective interactions to be important for some phenomena. Whereas the density dependence associated to the two-body correlations generated by the tensor force is thought to be included in ${\cal F}$, a better treatment of the tensor force in effective interactions, especially as a function of isospin, deserves additional work in the future.

The identification of $G_{LDA}^{\left(\alpha,\alpha\right)} (\vec{R}, \vec{r}, \vec{p})$ with $v_{Skyrme} (\vec{R}, \vec{r}, \overrightarrow{k}, \overleftarrow{k'})$ requires to neglect the tensor force in the bare interaction and to perform an expansion to second order in the range of the effective interaction $G_{LDA}^{\left(\beta,\alpha\right)} (\vec{R}, \vec{r}, \vec{p})$\footnote{The Gogny force also makes use of a zero-range density-dependent velocity-independent term.}; that is, for all powers $V^{l} (\vec{r}, \vec{p})$ of the bare interaction. Performing these expansions and grouping the terms coming from the central and the spin-orbit parts, one formally obtains:

\begin{eqnarray}
V^{l} (\vec{r}, \overrightarrow{k}, \overleftarrow{k'}) &\approx& \, V_{0} (l) \, \, \delta (\vec{r}) \nb \\
        && + \, V_{1} (l)  \,  \, (\delta (\vec{r}) \, \, \overrightarrow{k}^2 \, + \, \overleftarrow{k'}^2 \, \, \delta (\vec{r}) ) \nb  \\
        && + \, V_{2} (l) \, \, \overleftarrow{k'} . \, \, \delta (\vec{r}) \, \, \overrightarrow{k} \label{rangeexp}  \\
        && + \, \, V_{3} (l)  \, \, (\hat{\vec{\sigma_1}} \, + \, \hat{\vec{\sigma_2}} ) \, \, \overleftarrow{k'} \wedge \, \, \delta (\vec{r}) \, \, \overrightarrow{k}  \, \, \, \, . \nb
\end{eqnarray}

In this equation, the coefficients $V_{i} (l)$ incorporate dependences on $\vec{k}$ originally denoted as $\vec{p}$, but also on $\hat{\vec{\sigma}}_{1}$, $\hat{\vec{\sigma}}_{2}$ and $\hat{\vec{\tau}}_{1}.\hat{\vec{\tau}}_{2}$ originating from $V$. From a general point of view, it is unfortunately not possible for an given interaction to express these coefficients as a function of $l$ or through iterative relations. 

Here in fact, only the terms independent of the velocities should present a density-dependence in $G_{LDA}^{\left(\alpha,\alpha\right)} (\vec{R}, \vec{r}, \vec{p})$ in order to recover a Skyrme type interaction. Thus, the expansion~\ref{rangeexp} as to be cut to zero-order in $\overrightarrow{k}$ and $\overleftarrow{k'}$ for $l \ge 2$. A few coefficients only remain and one obtains:

\begin{eqnarray}
G_{LDA}^{\left(\beta,\alpha\right)} (\vec{R}, \vec{r}, \overrightarrow{k}, \overleftarrow{k'}) \, &\approx& \, V_{0} (1) \, \, \delta (\vec{r}) \nb  \\
        && + \, V_{1} (1)  \,  \, (\delta (\vec{r}) \, \, \overrightarrow{k}^2 \, + \, \overleftarrow{k'}^2 \, \, \delta (\vec{r}) ) \nb \\
        && + \, V_{2} (1) \, \, \overleftarrow{k'} . \, \, \delta (\vec{r}) \, \, \overrightarrow{k}  \label{expansionrange} \\
        && + \, \, V_{3} (1)  \, \, (\hat{\vec{\sigma_1}} \, + \, \hat{\vec{\sigma_2}} ) \, \, \overleftarrow{k'} \wedge \, \, \delta (\vec{r}) \, \, \overrightarrow{k} \nb \\
        && +  \, \sum_{l=2}^{\infty} \, V_{0} (l) \, \frac{{\cal F}^{l} \left[\rho^{\beta}_{0}(\vec{R})\right] \, - \, {\cal F}^{l} \left[\rho^{\alpha}_{0}(\vec{R})\right]}{{\cal F} \left[\rho^{\beta}_{0}(\vec{R})\right] \, - \, {\cal F} \left[\rho^{\alpha}_{0}(\vec{R})\right]}  \, \, \delta (\vec{r}) \, \, \, \, \, \, . \nb
\end{eqnarray}

\subsection{Mean-Field.}
\label{firstlevel0}

In order to really identify $G_{LDA}^{\left(\beta,\alpha\right)} (\vec{R}, \vec{r}, \overrightarrow{k}, \overleftarrow{k'})$ with a Skyrme force used at the mean-field level, one has to do $\alpha = \beta$ in Eq.~\ref{expansionrange}. Doing so, the identification can be done by truncating the power series appearing in $G_{LDA}^{\left(\alpha,\alpha\right)}$ to first order in ${\cal F}$, by taking:

\begin{equation}
{\cal F} \left[\rho_{0}^{\alpha}(\vec{R})\right] \, = \, a \, \left(\rho_{0}^{\alpha}(\vec{R})\right)^{\sigma} \, \, \, \, \, ,
\label{q}
\end{equation}
where $a$ is a constant, and by establishing the formal correspondance:

\begin{eqnarray}
&\begin{array}{lcl}
t_0 \, ( 1 + x_0 P_{\sigma} ) \, &\Leftrightarrow& \, V_{0} (1)    \\
\frac{1}{2} \, t_1 \, ( 1 +  x_1 P_{\sigma} )  \, &\Leftrightarrow& \,  V_{1} (1)  \\
t_2 \, ( 1 + x_2 P_{\sigma} ) \, &\Leftrightarrow& \,   V_{2} (1)   \\
\frac{1}{6} \, t_3 \, ( 1 + x_3 P_{\sigma} )  \, &\Leftrightarrow& \,  2 \, a  \, V_{0} (2)   \\
i W_0  \, &\Leftrightarrow& \,  V_{3} (1)   \, \, \, \, \, . 
\end{array}& \label{coeff1} 
\end{eqnarray}

As we were not able to derive ${\cal F}$ explicitely, Eq.~\ref{q} is to be understood as its phenomenological determination relying on existing successful effective interactions. The actual value of $\sigma$ has been a subject of debate. On one hand, general arguments were in favor of $\sigma = $ 2/3~\cite{bethe2}. Fits to diagonal Brueckner matrix elements at the Fermi level calculated from realistic forces gave $\sigma = $ 1/6~\cite{sprung} or 1/3~\cite{coon} as good values for the short-range repulsive as well as for the long-range attractive parts of the interaction; this being true for several spin/isospin channels. Similar calculations provided two different values for the short and the long-range parts of the interaction, namely 1 and 1/3~\cite{nege,siemens}. On the other hand, fits of phenomenological forces on empirical nuclear matter properties gave several values ranging from $\sigma = $ 1 for previous versions of the Skyrme force~\cite{beiner} to $\sigma = $ 1/6 for some recent versions~\cite{chab}. This low value has been necessary to get realistic values for both the effective mass $m^{\ast}_{\infty}$ and the compressibility $K_{\infty}$ of infinite symmetric nuclear matter. Such fits take into account the density dependence stemming from several physical effects as listed in section~\ref{background}.

\subsection{Beyond the Mean-Field.}
\label{firstlevel1}

We can now extend the density dependence of the effective interaction for calculations beyond the mean-field approximation. The hypothesis of a truncation to first order in ${\cal F}$ together with the previous phenomenological determination of this function provide the non-diagonal term of Eq.~\ref{expansionrange} ($\alpha \neq \beta$):

\begin{equation}
\frac{1}{6} \, \, t_3 \, \, ( 1 + x_3 P_{\sigma} ) \, \, \frac{[\rho_{0}^{\beta}(\vec{R})]^{\sigma} \, +  \, [\rho_{0}^{\alpha}(\vec{R})]^{\sigma}}{2} \, \,  \delta (\vec{r})  \, \, \, \, , \label{prescription}
\end{equation}
as the natural extension for the density-dependent part of the Skyrme force to be used in $\langle \, \Phi_{0}^{\beta} \, | \, \ldots \, | \, \Phi_{0}^{\alpha} \rangle$. The derived prescription is different from those used until now in GCM or projected mean-field calculations. For projection on particle number, one can check that Eq.~\ref{prescription} reduces to a dependence on a single local mean-field density since the mean-field local density $\rho^{0}_{0} (\vec{R})$ is invariant under rotation in gauge space. 

Eq.~\ref{prescription} constitutes the main result of the present section. It has been obtained through a sequence of reasonable approximations. The last ones have been performed in relation with the specific form of the Skyrme force. The presently developed scheme is more general and one could avoid some approximations in order to generalize Eq.~\ref{prescription}. For instance, one could keep a finite-range density-dependent term or at least the density-dependence of the velocity-dependent terms in Eq.~\ref{expansionrange} as suggested in Ref.~\cite{coon}. One could also consider higher powers of ${\cal F}$ in Eq.~\ref{ldagbetaalpha} or in Eq.~\ref{expansionrange}. The reduction of the original isospin-dependent densities to the isoscalar density could be avoided, at least for exotic nuclei.The reduction of the original signature-dependent densities to the scalar density could be avoided for odd-even and rotating nuclei\footnote{Then, one should be careful to end up with an energy functional invariant under time-reversal symmetry.}. Last but not least, the dependence on starting energies should also be kept since it is known to play a role~\cite{kohl}. However, all these extensions would correspond to more general forms of the Skyrme or Gogny forces at the mean-field level which have not been considered up to now in quantitative microscopic calculations of finite nuclei. Following the strategy behind the use of phenomenological forces, implementing such further complications in the interaction will have to be motivated by clear experimental hints.

\subsection{Higher-Orders.}
\label{qpcase}

The generalized Skyrme force derived from the $G^{\left(\beta,\alpha\right)}$ interaction on the energy shell should not be used beyond the lowest order in the perturbative expansion of the energy. However, it can be useful in some systems to include diabatic effects in the GCM for instance~\cite{tajima1}. Associated diagrams correspond to non-zero orders in the extended Goldstone-Brueck\-ner perturbation theory~\cite{duguet}. They make use of $G^{\left(\beta,\alpha\right)}$ off the energy shell for which the energy denominator appearing in Eq.~\ref{projecteur2} is modified. Within the local approximation, the functional ${\cal F}$ should be changed for each vertex with respect to its definition at the lowest order. This means that the use of Eq.~\ref{prescription} in the GCM is questionable when including diabatic effects. This deserves additional work in the future.

\section{Multi-Body Forces Renormalization.}
\label{multibody}

We now turn to another origin of the density dependence of the two-body effective interaction. Let us make the hypothesis that the actual Hamiltonian of the $N$-body nuclear system reads as:

\begin{equation}
H^{(3)} \, = \, \sum_{i,j} \,  t_{ij} \, c^{\dagger}_i \, c_j  \, +  \,  \frac{1}{4} \,  \sum_{i,j,k,l} \, \overline{V}^{(2)}_{iljk} \, \,  c^{\dagger}_i \, c^{\dagger}_l \, c_k \, c_j \, + \,  \frac{1}{36} \,  \sum_{i,j,k,l,m,n} \, \overline{V}^{(3)}_{ilnjkm} \, \,  c^{\dagger}_i \, c^{\dagger}_l \, c^{\dagger}_n \, c_m \, c_k \, c_j \, \, \, ,  \, \, \, \, \label{H3body}
\end{equation}
where $\bar{V}^{(2)}_{iljk}$ and $\bar{V}^{(3)}_{ilnjkm}$ are antisymmetrized matrix elements. The creation and annihilation operators $\{c^{\dagger}_{i},c_{j}\}$ refer to the single-particle basis $\{\varphi_{i}\}$. We consider at this stage that two-body correlations have already been renormalized in such a way that $V^{(2)}$ is to be seen as the Brueckner matrix $G_{LDA}^{\left(\beta,\alpha\right)}$ and that the three-body vertex approximately includes these correlations through:

\begin{equation}
V^{(3)} (\vec{r_1},\vec{r_2},\vec{r_3}) \, \equiv \, \tilde{V}^{(3)} (\vec{r_1},\vec{r_2},\vec{r_3}) \, (1-\eta(\vec{r_1}-\vec{r_2})) \, (1-\eta(\vec{r_2}-\vec{r_3})) \, (1-\eta(\vec{r_3}-\vec{r_1})) \, \, ,  \, \, \, \, \label{2bodycorrelin3body}
\end{equation}
where $\tilde{V}^{(3)}$ is the actual bare three-body force and $\eta(\vec{r_1}-\vec{r_2})$ is some average of the two-body defect wave function over occupied states~\cite{day}. The proper treatment of these correlations together with a three-body force relies on the Bethe-Faddeev equations~\cite{day2}.

In what follows, no density dependence appear in $V^{(2)}$ and $V^{(3)}$, unless otherwise specified. Of course, the previous statement about two-body correlations implies that some density dependences originating from two-body correlations are contained in the first place in both the effective two-body and three-body interactions. We will return to this issue later.

We separate the GCM from the projected mean-field method since the energy minimization is performed with respect to different variational parameters in the two cases.

\subsection{GCM and Three-Body Force.}
\label{gcm}

In order to identify the density dependence accounting for three-body force effects, we calculate and minimize the energy for two different Hamiltonians. First, the three-body force is taken into account but no density-dependence occurs in the two-body one. Then, the three-body force is omitted in the Hamiltonian but a density dependence is introduced explicitly in the two-body force. In this second case the Hamiltonian is denoted as $H^{(3)}_{eff}$.

$\bullet$ We obtain in the first case using the general Wick theorem~\cite{bal1}:

\begin{eqnarray}
\langle \, \Psi_{k} \, | \, H^{(3)} \, | \, \Psi_{k} \rangle \, &=& \, \sum_{\beta,\alpha} \, f^{k \, \ast}_{\beta} f^{k}_{\alpha} \, \langle \, \Phi^{\beta}_{0} \, | \, H^{(3)} \, | \, \Phi^{\alpha}_{0}  \rangle \label{eneragain} \\
&& \nb \\
&=& \, \sum_{\beta,\alpha} \, f^{k \, \ast}_{\beta} f^{k}_{\alpha} \, \left[  \sum_{i,j} \,  t_{ij} \, \rho^{\left(\beta,\alpha\right)}_{ji} \, \, +  \, \, \frac{1}{2} \,  \sum_{i,j,k,l} \, \overline{V}^{(2)}_{iljk} \,  \rho^{\left(\beta,\alpha\right)}_{ji} \, \rho^{\left(\beta,\alpha\right)}_{kl} \right.  \nb \\
&& \label{ener3body}  \\
&& \, \, \,  + \, \,\left.  \frac{1}{6} \,  \sum_{i,j,k,l,m,n} \, \overline{V}^{(3)}_{ilnjkm} \,  \rho^{\left(\beta,\alpha\right)}_{ji} \, \rho^{\left(\beta,\alpha\right)}_{kl} \,  \rho^{\left(\beta,\alpha\right)}_{mn} \, \right] \,  \langle \, \Phi^{\beta}_{0} \, | \, \Phi^{\alpha}_{0} \rangle \, \, \, \, \, \, , \nb
\end{eqnarray}
where

\begin{equation}
\rho^{\left(\beta,\alpha\right)}_{ji} \, = \, \frac{\langle \, \Phi_{0}^{\beta} \, | \, c^{\dagger}_{i} \, c_{j} \, | \, \Phi_{0}^{\alpha} \rangle}{\langle \, \Phi_{0}^{\beta} \, | \, \Phi_{0}^{\alpha} \rangle} \, \, \, ,  \, \, \, \, \label{densmixte}
\end{equation}
denotes a matrix element of the mixed one-body density operator. The minimization of the  energy with respect to the $f^{k \, \ast}_{\beta}$ gives:

\begin{eqnarray}
\sum_{\beta} \frac{\delta}{\delta f^{k \, \ast}_{\beta}} \, \left[ \frac{\langle \, \Psi_{k} \, | \, H^{(3)} \, | \, \Psi_{k} \rangle}{\langle \, \Psi_{k} \, | \, \Psi_{k} \rangle} \right] \, \delta f^{k \, \ast}_{\beta} \, &=& \, 0  \, \, \, \, \, \, ,
\label{mini3}
\end{eqnarray}
for all $\delta f^{k \, \ast}_{\beta}$, which can be recast into a set of coupled equations of motion:

\begin{eqnarray}
&& \sum_{\alpha} \, f^{k}_{\alpha} \, \left[  \sum_{i,j} \,  t_{ij} \, \rho^{\left(\beta,\alpha\right)}_{ji} \, \, +  \, \, \frac{1}{2} \,  \sum_{i,j,k,l} \, \overline{V}^{(2)}_{iljk} \,  \rho^{\left(\beta,\alpha\right)}_{ji} \, \rho^{\left(\beta,\alpha\right)}_{kl} \right. \nb \\
&& \label{equamotion3} \\
&& \hspace{3cm} + \, \,\left.  \frac{1}{6} \,  \sum_{i,j,k,l,m,n} \, \overline{V}^{(3)}_{ilnjkm} \,  \rho^{\left(\beta,\alpha\right)}_{ji} \, \rho^{\left(\beta,\alpha\right)}_{kl} \,  \rho^{\left(\beta,\alpha\right)}_{mn} \, \right]   \langle \, \Phi^{\beta}_{0}  | \, \Phi^{\alpha}_{0} \rangle  \nb \\
&& \nb \\
&& \hspace{8cm}= \,  {\cal E}_{k}^{mix} \,  \sum_{\alpha} \, f^{k}_{\alpha} \,  \langle \, \Phi^{\beta}_{0}  | \, \Phi^{\alpha}_{0} \rangle \, \, \, \,  ,  \, \, \, \, \, \, \, \, \nb
\end{eqnarray}
for all $\beta$.

$\bullet$ Omitting the three-body force, we proceed to the same calculation using an effective two-body force depending linearly on the mixed density\footnote{For simplicity, we do not write the dependences of $V^{(3)(\beta,\alpha)}_{eff}$, $V^{(2)}$ and $v$ on the relative momentum $\vec{p}$, the spin and isospin operators $(\hat{\vec{\sigma}}_{1}, \hat{\vec{\sigma}}_{2}, \hat{\vec{\tau}}_{1}.\hat{\vec{\tau}}_{2})$.}:

\begin{equation}
V^{(3)(\beta,\alpha)}_{eff} (\vec{r},\vec{R}) \, = \, V^{(2)} (\vec{r}) \, + \, v (\vec{r}) \, \rho^{\left(\beta,\alpha\right)} (\vec{r}_{1}, \vec{r}_{2})  \, \label{defVeff2} \, \, \, \, \, \, ,
\end{equation}
where $\vec{r}$ and $\vec{R}$ are respectively the relative and center of mass position vectors of the two interacting nucleons while $\rho^{\left(\beta,\alpha\right)} (\vec{r}_{1}, \vec{r}_{2})$ denotes the non-local mixed nucleon density:

\begin{eqnarray}
\rho^{\left(\beta,\alpha\right)} (\vec{r}_{1}, \vec{r}_{2})  \, &=& \, \sum_{s_{z},s'_{z},t,t'} \, \rho^{\left(\beta,\alpha\right)} (\vec{r}_{1} \, s_{z} \, t, \, \vec{r}_{2} \, s'_{z} \, t') \nb \\
&& \nb \\
&=& \, \sum_{ij} \, \varphi_{I}^{\ast} (\vec{r}_{2}, \zeta_{z}', s'_{z}, t) \, \varphi_{J} (\vec{r}_{1}, \zeta_{z}, s_{z}, t) \, \rho^{\left(\beta,\alpha\right)}_{ji} \label{c11} \\
&& \nb \\
&=& \, \rho^{\left(\beta,\alpha\right)}_{0} (\vec{r}_{1}, \vec{r}_{2}) \, + \, (s^{\left(\beta,\alpha\right)}_{0})_{x} (\vec{r}_{1}, \vec{r}_{2}) \nb \, \, \, \, \, .
\end{eqnarray}
where $\rho^{\left(\beta,\alpha\right)}_{0}$ and $\vec{s}^{\, \left(\beta,\alpha\right)}_{0}$ are its scalar-isoscalar and vector-isoscalar parts~\cite{engel}. The non appearence of isovector components in $\rho^{\left(\beta,\alpha\right)} (\vec{r}_{1}, \vec{r}_{2})$ is due to the fact that we restrict our study to systems where protons and neutrons are not mixed. The interaction defined through Eq.~\ref{defVeff2} depends on $(\beta,\alpha)$ and is to be used in the corresponding matrix element $\langle \, \Phi^{\beta}_{0} \, | H^{(3) (\beta,\alpha)}_{eff} |  \, \Phi^{\alpha}_{0} \rangle$. Thus, one gets:

\begin{equation}
\langle \, \Psi_{k} \, |  H^{(3)}_{eff}  | \, \Psi_{k} \rangle\!\equiv\!\sum_{\beta,\alpha} \, f^{k \, \ast}_{\beta} f^{k}_{\alpha}  \left[  \sum_{i,j} \,  t_{ij} \, \rho^{\left(\beta,\alpha\right)}_{ji} \, +  \, \frac{1}{2} \,  \sum_{i,j,k,l} \, \left(\overline{V}^{(3)(\beta,\alpha)}_{eff}\right)_{iljk} \,  \rho^{\left(\beta,\alpha\right)}_{ji} \, \rho^{\left(\beta,\alpha\right)}_{kl}  \right]   \langle \, \Phi^{\beta}_{0}  | \, \Phi^{\alpha}_{0}  \rangle  \, \, \, \, \, \, \, \,  . \label{ener2body}
\end{equation}

Developing $\bar{V}^{(3)(r,s)}_{eff}$ in Eq.~\ref{ener2body}, one obtains the same expression as that given by Eq.~\ref{ener3body} with $\bar{V}^{(3)}_{ilnjkm}$ replaced by the matrix element $v_{ilnjkm}$ defining an effective three-body force:

\begin{eqnarray}
v_{ilnjkm} &=& 3 \int\!\int d\vec{r}_{1} \, \, d\vec{r}_{2} \, \varphi_{i}^{\ast} (\vec{r}_{1}) \, \varphi_{l}^{\ast} (\vec{r}_{2}) \, \varphi_{n}^{\ast} (\vec{r}_{2}) \, v (\vec{r}_{1}-\vec{r}_{2}, \vec{p}, \hat{\vec{\sigma}}_{1}, \hat{\vec{\sigma}}_{2}, \hat{\vec{\tau}}_{1}.\hat{\vec{\tau}}_{2}) \, \nb  \\
&& \label{matrixelementeff} \\
&& \hspace{3.7cm} \, \varphi_{m} (\vec{r}_{1}) \, \left[ \varphi_{k} (\vec{r}_{1}) \varphi_{j} (\vec{r}_{2}) \, - \, \varphi_{j} (\vec{r}_{1}) \varphi_{k} (\vec{r}_{2}) \right] \, \, \, \, \, \, \, \, . \nb
\end{eqnarray}

Thus, the correlated energy in the state $| \, \Psi_{k} \rangle$ for the Hamiltonian $H^{(3)}$ is finally reproduced term by term for all spin/isospin indices. The key point of this derivation is the {\it mixed} nature of the density inserted in the effective two-body force.

The equivalence is not complete since the effective matrix elements~\ref{matrixelementeff} cannot simulate all antisymmetrized matrix elements of an arbitrary three-body force $V^{(3)} (\vec{r_1},\vec{r_2},\vec{r_3})$, in particular because of their non-antisymmetrized character in $(k,m)$ and $(j,m)$. However, the freedom in the choice of the two-body term $v (\vec{r}, \vec{p}, \hat{\vec{\sigma}}_{1}, \hat{\vec{\sigma}}_{2}, \hat{\vec{\tau}}_{1}.\hat{\vec{\tau}}_{2})$ can be used to make $v_{ilnjkm}$ reproduce $\overline{V}^{(3)}_{ilnjkm}$ as accurately as possible. The possibility of an exact equivalence has already been discussed at the mean-field level in connection with particular forms of the three-body effective interaction. In practice, this has only been done for zero-range forces. It has been shown how a Skyrme force depending linearly on the mean-field matter density in the (S=1,T=0) channel allows for the reproduction of the HF energy obtained with a zero-range three-body force in a spin-saturated (time-reversal invariant) system~\cite{vauth}. Alternatively, by changing the spin-isospin dependence of the density-dependent two-body term,  Onishi and Negele were able to reproduce the effect of the zero range three-body force for the HF energy, single-particle spectrum and two-body p-h matrix elements of a spin-{\it isospin} saturated system~\cite{onish}. Both versions could be easily recovered by taking $\alpha=\beta$ in previous expressions; the local part of the density $\rho^{\alpha}_{0} (\vec{R})$ being selected by the zero-range character of the force. In both cases however, the equivalence came out to be invalid for systems breaking time-reversal symmetry. It also became clear that the use of a simple zero-range three-body force having a necessary repulsive nature in the (S=1,T=0) channel led to spin instabilities in time-reversal invariant systems~\cite{chang,back,string}. The exact equivalence in systems breaking time reversal symmetry is not achievable if one does not keep any spin-density dependence in the zero-range two-body term, as shown by Eq.~\ref{c11}. However, the spin instability associated to the Skyrme interaction can be cured by restricting oneself to a density-dependent two-body term with appropriate spin-isospin dependence~\cite{chang}, by keeping a finite-range three-body force or at least by considering attractive velocity-dependent three-body terms. The last option however often induces a collapse in the equation of state of symmetric nuclear matter at high density~\cite{back,waroquier} while a finite range would destroy the numerical convenience of the Skyrme interaction.

The reduction of the density dependence of the two-body term to the scalar part of the density matrix is a stronger limitation beyond than at the mean field level. Indeed, even for time reversal invariant systems, the vector part of the local mixed density is non zero for $\alpha \neq \beta$~\cite{bonche3}. If one forgets about the dependence on $\vec{s}^{\, \left(\beta,\alpha\right)}_{0}$ in Eq.~\ref{defVeff2}, non-diagonal matrix elements $v_{ilnjkm}$ in $(n_{s_{z}},m_{s_{z}})$ will not be considered. By keeping this dependence, one would have to be careful about spin instabilities as discussed before. 

The above derivation is a general starting point for the quantitative renormalization of three-body forces. First, it shows the mixed nature of the density required for non-diagonal $N$-body matrix elements. Second, we think that it will be useful to move ahead towards the next step which has to be the gross renormalization of {\it realistic} three-body interactions. Indeed, rather than the reproduction of a simplified analytical three-body interaction, which is itself phenomenological, it would be more worthwhile to reproduce the main properties of a three-body force derived from an underlying field theory. Using Eq.~\ref{matrixelementeff}, together with a sufficiently simple two-body effective interaction which satisfies Eq.~\ref{c11}, could help to do so. In particular, the repulsive or attractive character of three-body forces derived from microscopic meson-exchange models has been characterized in each (S,T) channel in nuclear matter calculations~\cite{grange,zuo}. The subtle combination of these contributions as a function of the density is an important part of the saturation process. It allows a correct reproduction of the empirical values of the density and energy per particle at the saturation point in Brueckner-Hartree-Fock calculations~\cite{zuo}. These combined contributions are also important to describe asymmetric nuclear matter correctly as a function of isospin. Consequently, the different channels of the three-body force should be treated carefully in mean-field type calculations using phenomenological effective forces. In particular, one has to reconcile the crucial binding effect of three-body forces in light nuclei~\cite{wiringa} with its saturation character at normal density of nuclear matter~\cite{grange,zuo}.

Varying the energy given by Eq.~\ref{ener2body} with respect to $f^{k \, \ast}_{\beta}$, the same equations of motion as with the Hamiltonian $H^{(3)}$ are obtained. The choice of the mixed density in $V^{(3)(\beta,\alpha)}_{eff} (\vec{r},\vec{R})$ leads to a zero rearrangement term:

\begin{equation}
\Gamma_{\delta V} \, = \, \sum_{\beta,\alpha} \, f^{k \, \ast}_{\beta} f^{k}_{\alpha} \, \langle \, \Phi^{\beta}_{0}  | \,  \frac{\partial V^{(3)(\beta,\alpha)}_{eff} (\vec{r},\vec{R})}{\partial \rho^{\left(\beta,\alpha\right)} (\vec{r}_{1},\vec{r}_{2})} \,\frac{\partial \rho^{\left(\beta,\alpha\right)} (\vec{r}_{1},\vec{r}_{2})}{\partial f^{k \, \ast}_{\beta}} \, | \, \Phi^{\alpha}_{0}  \rangle \, / \, \langle \, \Psi_{k} \, | \, \Psi_{k} \rangle \, = \, 0 \, \, \, \, \, \, \, \, ,
\label{rearrangement1}
\end{equation}
since $\rho^{\left(\beta,\alpha\right)} (\vec{r}_{1},\vec{r}_{2})$ is independent of the mixing coefficients $f^{k \, \ast}_{\beta}$.

Thus, the contribution from the three-body force to Eq.~\ref{equamotion3} is recovered from the redefinition of the two-body force only. As demonstrated in appendix~\ref{calcgcmmixing}, the use of the correlated density would generate redundant terms, notably through a non-zero rearrangement term.

\subsection{Projection and Three-Body Force.}
\label{proj}

The same question as in the previous section is addressed for projection type configuration mixings. In this case, the VAP is performed with respect to the individual wave functions defining the product state from which $| \, \Psi_{k}  \rangle$ is projected out. As a application, we consider the restoration of angular momentum $I$, with projection $M$, for an axially symmetric Slater-determinant. The projected state reads as:

\begin{eqnarray}
| \, \Psi_{IM}  \rangle \, &=& \, \sum_{\alpha= -n}^{n} \, f^{IM}_{\alpha}  \, | \, \Phi^{\alpha}_{0} \rangle \nb \\
&& \label{mixingproj} \\
&=& \, \hat{P}_{IM} \, | \, \Phi^{\alpha}_{0} \rangle \, \, \, \, , \nb
\end{eqnarray}
where $\hat{P}_{IM}$ is the angular momentum projector~\cite{yocco} and $| \, \Phi^{\alpha}_{0}(IM) \rangle$ is written in terms of a fixed product state of reference $| \, \Phi  \rangle$ through:

\begin{equation}
| \, \Phi^{\alpha}_{0} \rangle \, = \, e^{\frac{1}{2} \sum_{u,u'} Z^{IM}_{uu'} \, c^{\dagger}_{u} \, c_{u'}} \, | \, \Phi  \rangle \, \, \, \, \, \, \, \, .
\label{parametrization}
\end{equation}

Note that $| \, \Phi^{\alpha}_{0} \rangle$ implicitely depends on $(IM)$ since the minimization procedure provides a different product state for each value of these quantum numbers. 

$\bullet$ The variation of the mean-energy in the state $| \, \Psi_{IM}  \rangle$ is done with respect to the $Z^{IM}_{uu'}$ and reads as:

\begin{eqnarray}
\delta \, \frac{\langle \, \Psi_{IM} \, |  H^{(3)} | \, \Psi_{IM} \rangle}{\langle \, \Psi_{IM}  | \, \Psi_{IM} \rangle} &=& \frac{1}{2} \, \, \sum_{u,u'}^{} \left[\frac{\langle \, \Psi_{IM} \, | \, c^{\dagger}_{u} \, c_{u'} \,  H^{(3)} \, | \, \Psi_{IM} \rangle}{\langle \, \Psi_{IM}  | \, \Psi_{IM} \rangle} \right. \nb \\
&&  \nb \\
&&\left. \hspace{1.5cm} - \, \, \frac{\langle \, \Psi_{IM} \, | \, c^{\dagger}_{u} \, c_{u'} \, | \, \Psi_{IM} \rangle \,\langle \, \Psi_{IM} \, |  \, H^{(3)} \, | \, \Psi_{IM} \rangle}{\langle \, \Psi_{IM}  | \, \Psi_{IM} \rangle^2}\right]  \delta Z^{IM}_{uu'}  \nb \\
&&  \nb \\
&=& \,  0 \, \, \, \ , \label{miniproj}
\end{eqnarray}
for all $\delta Z^{IM}_{uu'}$. The equations of motion are:

\begin{equation}
\langle \, \Psi_{IM} \, | \, c^{\dagger}_{u} \, c_{u'} \,  H^{(3)} \, | \, \Psi_{IM} \rangle \, \, = \, \, {\cal E}_{IM}^{mix} \, \langle \, \Psi_{IM} \, | \, c^{\dagger}_{u} \, c_{u'} \, | \, \Psi_{IM} \rangle \, \, \, \, ,
\label{equamotionproj}
\end{equation}
for all couples $(u,u')$. These equations are valid only if no rearrangement term appears. This is the case with $H^{(3)}$. Eq.~\ref{equamotionproj} expanded in terms of mixed densities is given in appendix~\ref{appequamotion}.

$\bullet$ As for the GCM, we do the calculation using the Hamiltonian $H^{(3)(\beta,\alpha)}_{eff}$. In this case, there is a non-zero rearrangement term in the equations of motion. The calculation is performed in appendix~\ref{reaterm}. The comparison between the contribution arising from the three-body force and those coming from the redefinition of the two-body force and the rearrangement term is also proposed. This calculation shows that the equivalence between the two is once again obtained thanks to the choice of the mixed density in the effective interaction.

\subsection{Generalization to Multi-Body Forces.}
\label{generalisation}

The formal equivalence between a three-body force and a two-body one depending linearly of the mixed density has been shown in the context of configuration mixing calculations. However, most of nowadays phenomenological interactions depend on the density through a non-linear function $\rho^{\sigma}$ with $0 < \sigma \leq 1$ (e.g. 1/6 for the Skyrme force SLy4~\cite{chab}). Even if such a dependence certainly accounts for several physical effects, we can interpret it as coming from the renormalization of multi-body forces effects. In order to do that, $\rho^{\sigma}$ is written in terms of a power series around the nuclear saturation density $\rho_{sat}$:

\begin{eqnarray}
\rho^{\sigma} \, &=& \, \rho_{sat}^{\sigma} \, \sum_{n} \, a^{(\sigma)}_n \, \left(\frac{\rho - \rho_{sat}}{\rho_{sat}}\right)^{n} \, \, \, \, ,  \label{devexact}\\
&& \nb \\
&\approx& \, \rho_{sat}^{\sigma} \, \sum_{k=0}^{K} \, b^{(\sigma)}_k \, \left(\frac{\rho}{\rho_{sat}}\right)^{k} \, \, \, \, .
\label{devapprox}
\end{eqnarray}

The domain of validity of the expansion~\ref{devexact} is $]0,2\rho_{sat}[$. We rearrange it as a function of the successive integer powers of $\rho$. Doing so, each coefficient of $\rho^{k}$ is in principle divergent as no expansion exists for $\rho^{\sigma}$ around 0 when $\sigma \not\in \mathbb{N}$. Thus, we approximate Eq.~\ref{devexact} by cutting the sum at some order $n=K$. In this way, we obtain a good approximation of $\rho^{\sigma}$ on the domain $[\epsilon(K),\rho_{sat}]$\footnote{$\epsilon(K) \longrightarrow 0$ for $K \longrightarrow \infty$.} and can reorder the finite number of term as a function of $\rho^{k}$. This gives the formal expansion~\ref{devapprox}.

In the following, we use such an expansion to interpret the full density dependence $\rho^{\sigma}$ as coming from multi-body forces in the nuclear Hamiltonian $H^{(K)}$; the linear term of Eq.~\ref{devapprox} being related to the three-body force, the squared term to the four-body force etc\ldots Starting from such an hypothesis, one can show, using the same technique as in the previous sections for the three-body force, that the two-body force:

\begin{eqnarray}
V^{(K)(\beta,\alpha)}_{eff} (\vec{r},\vec{R}) \, &=&  \, V^{(2)} (\vec{r}) \, + \, v (\vec{r}) \, \rho_{sat}^{\sigma} \, \sum_{k=1}^{K-2} \, b^{(\sigma)}_k \, \left( \frac{\rho^{\left(\beta,\alpha\right)} (\vec{r}_{1}, \vec{r}_{2})}{\rho_{sat}} \right)^{k}  \, \label{defVeff2gene} ,
\end{eqnarray}
allows to formally reproduce the energy of a correlated state for a Hamiltonian having two, three, four, \ldots K-body forces. The non-antisymmetrized matrix elements $v^{("p")}_{il\ldots jk\ldots}$ of the corresponding effective ``p-body'' interaction are defined from the term with $k=p-2$ in Eq.~\ref{defVeff2gene} in the same spirit and with the same limitations as $v_{ilnjkm}$ in section~\ref{gcm}. Finally, this calculation motivates a term proportional to $[\rho^{\left(\beta,\alpha\right)}_{0} (\vec{R})]^{\sigma}$ as approximately renormalizing multi-body forces effects, the crucial point being again the use of the mixed density.

Finally, let us mention that identical calculations mixing non-orthogonal HFB quasi-particle states instead of Slater determinants would have led to the same conclusion for the density-dependence induced in the particle-particle channel as soon as the terms up to second order only in the pairing tensor are kept in the energy.

\subsection{Quasi-Particle Type Configuration Mixing.}
\label{quasi-particleconfig}

Two different cases occur when a mixing of individual excitations is considered to include small amplitude correlations and diabatic effects in the trial state. 

First, each particle-hole state $| \, \Phi^{\alpha}_{i}  \rangle$ is calculated self-consistently through the minimization of its energy. In this case, the $| \, \Phi^{\alpha}_{i}  \rangle$ are fixed non-orthogonal product wave-functions and the variation is performed with respect to $f^{k \, \ast}_{i}$. The situation is identical to the GCM and the same conclusions as regard the renormalization of multi-body forces effects hold.

Second, particle-hole states are calculated perturbatively with respect to each ground-state Slater-determinant $| \, \Phi^{\alpha}_{0}  \rangle$. In this case the individual excitations $| \, \Phi^{\alpha}_{i}  \rangle$ referring to a given vacuum are orthogonal and the above calculations do not hold since the generalized Wick theorem cannot be used. For such a configuration mixing, and whatever the variational parameters are, we were not able to obtain any formal equivalence between a particular two-body density-dependent interaction and a three-body one.

\section{Skyrme Force Beyond the Mean-Field.}
\label{appphenforces2}

Given the results obtained in the previous sections, we propose a simple extension of the Skyrme force for configuration mixing calculations such as the GCM and the projected mean-field method:

\begin{eqnarray}
v_{Skyrme}^{\left(\beta,\alpha\right)} (\vec{R}, \vec{r}, \overrightarrow{k}, \overleftarrow{k'}) \, &=& \, t_0 \, ( 1 + x_0 P_{\sigma} ) \, \, \delta (\vec{r}) \nb  \\
        && + \, \frac{1}{2} \, \, t_1 \, \,  ( 1 +  x_1 P_{\sigma} )  \, \, (\delta (\vec{r}) \, \, \overrightarrow{k}^2 \, + \, \overleftarrow{k'}^2 \, \, \delta (\vec{r}) ) \nb \\
        && + \, t_2 \, \,  ( 1 +  x_2 P_{\sigma} )  \, \, \overleftarrow{k'} . \, \, \delta (\vec{r}) \, \, \overrightarrow{k}  \label{prescription2} \\
        && + \, \frac{1}{6} \, \, X \, t_3 \, \, ( 1 + x_3 P_{\sigma} ) \, \,  \frac{[\rho_{0}^{\beta}(\vec{R})]^{\sigma} \, +  \, [\rho_{0}^{\alpha}(\vec{R})]^{\sigma}}{2}  \, \,  \delta (\vec{r})  \nb \\
&& + \, \frac{1}{6} \, \, (1 - X) \, t_3 \, \, ( 1 + x_3 P_{\sigma} ) \, \, [\rho^{\left(\beta,\alpha\right)}_{0} (\vec{R})]^{\sigma}   \, \,  \delta (\vec{r})  \nb   \\
        && + \, \, i W_0  \, \, (\hat{\vec{\sigma_1}} \, + \, \hat{\vec{\sigma_2}} ) \, \, \overleftarrow{k'} \wedge \,  \, \delta (\vec{r}) \,  \, \overrightarrow{k}  \, \, \, \, . \nb
\end{eqnarray}

In Eq.~\ref{prescription2},  $X$ is an adjustable parameter expressing our lack of knowledge about the relative weight of the two renormalized effects. The two types of densities used coincide with the standard local mean-field density when going back to a diagonal $N$-body matrix element. As they are considered with the same exponent $\sigma$, $v_{Skyrme}^{\left(\beta,\alpha\right)}$ reduces to the usual Skyrme force when going back to the mean-field level. In other words, the proposed Skyrme interaction is the simplest theoretically motivated extension for the calculation of non-diagonal $N$-body matrix elements. Note that, since two-body correlations are taken into account in Eq.~\ref{2bodycorrelin3body}, the term proportional to $(1-X)$ in Eq.~\ref{prescription2} should depend both on the mixed and the mean-field densities. However, ladders diagrams in connection with the three-body interaction do not play a significant role at and below the saturation density~\cite{zuo}. This is due to the small probability at low density for two nucleons to get very close to each other and feel the influence of a third in the same time. Not taking these correlations into account should not be a strong limitation for finite nuclei. Concerning the term proportional to $X$, its possible generalizations have already been discussed in section~\ref{firstlevel1}.

Given the different origins of the two density-dependent terms, taking identical analytical expressions for both is a restrictive and non motivated choice. In particular, it would be reasonable to have different exponents. The predominance of three-body over higher multi-body forces suggests an exponent close to one for the associated term whereas a smaller exponent would be appropriate to the resummation of two-body correlations. Similarly, the spin dependence $( 1 + x_3 P_{\sigma})$ should be different in the two terms. However, such a differentiation asks for a redefinition of the Skyrme interaction at the mean-field level. Such a work is underway~\cite{karim}. The essential to retain is that going beyond the mean-field approximation distinguishes the two origins of the density dependence by making two kinds of density appear and opens a new degree of freedom in the interaction.

\subsection{Extended Skyrme Functional.}
\label{functiobeyond}

Using the generalized Skyrme force~\ref{prescription2}, the approximate ground-state energy~\ref{energy} takes the form:

\begin{eqnarray}
{\cal E}^{mix}_{0}  \, \equiv \, \frac{\langle \, \Psi_{0} \, | \, H \, | \, \Psi_{0} \rangle}{\langle \, \Psi_{0} \, | \, \Psi_{0} \rangle} \, \, &\equiv& \frac{\sum_{\alpha,\beta} \, f_{\beta}^{0 \, \ast} f_{\alpha}^{0} \langle \, \Phi^{\beta}_{0} \, | \, H^{\left(\beta,\alpha\right)} \left[\rho_{0}^{\beta}(\vec{R}), \rho_{0}^{\alpha}(\vec{R}), \rho^{\left(\beta,\alpha\right)}_{0} (\vec{R})\right] \, | \, \Phi^{\alpha}_{0}  \rangle}{\sum_{\alpha,\beta} \, f_{\beta}^{0 \, \ast} f_{\alpha}^{0} \langle \, \Phi^{\beta}_{0} \, | \, \Phi^{\alpha}_{0}  \rangle} \nb \\ 
&&  \label{energylda} \\ 
&=& \, \sum_{\alpha,\beta} \, f_{\beta}^{0 \, \ast}  f_{\alpha}^{0} \, \int \, d\vec{R} \, \, \, {\cal H}^{\left(\beta,\alpha\right)} (\vec{R})  \, \, \, \, \, . \nb
\end{eqnarray}
where ${\cal H}^{\left(\beta,\alpha\right)} (\vec{R})$ is a functional of local densities only. It includes the local scalar-isoscalar mean-field and mixed densities originating from the generalized Skyrme force, and also the local mixed densities as coming from the non-diagonal contractions in Eq.~\ref{energylda}. Time-odd components of the force are always switched on in the context of mixed vacua. It makes non-zero time-odd local densities emerge in the energy functional such as the spin density, the current density or the vector part of the kinetic energy density. The explicit form of the Skyrme functional for configuration mixing calculations is given in Ref.~\cite{bonche3} and should be corrected in agreement with the newly derived density dependence of the effective interaction. 

The second equality in Eq.~\ref{energylda} is a matter of definition only since the r.h.s. cannot be re-factorized into the l.h.s., that is, as the mean value of a two-body operator in the state $| \, \Psi_{0} \rangle$. To make the meaningful effective interaction appear explicitly, the correlated energy had to be fully expanded in terms of the mixed product states. As discussed in Ref.~\cite{duguet}, this expresses the fact that the energy obtained from an effective force is more to be seen as a functional of (local) densities than as the mean value of a two-body Hamiltonian in a definite state. This is in fact already true at the mean-field level. Note however that the functional as considered here does not aim at renormalizing all correlations since it is defined at some precise order of a perturbative expansion. It keeps, at least formally, a link with the original bare force.

In order to obtain the energy as given by Eq.~\ref{energylda}, we have interpreted the Skyrme force as coming from an expansion in the range of the effective interaction. For Negele~\cite{nege2}, this force should rather be interpreted as a result of density matrices expansions in the energy density ${\cal H}^{\left(\beta,\alpha\right)} (\vec{R},\vec{r})$ obtained using a finite range $G_{LDA}^{\left(\beta,\alpha\right)}$ interaction as given by Eq.~\ref{ldagbetaalpha}. In this context, the Skyrme force would result from an additional average over the occupied states and its parameters would contain a combination of informations about both the initial effective two-body interaction $G^{\left(\beta,\alpha\right)}$ and two Fermi seas. Within this interpretation, the extended Skyrme force would provide the energy functional with coefficients themselves depending on local mixed density-matrices.

\section{Application: Symmetry Restoration.}
\label{prescriptionproj}

It is worth illustrating the previous result in the particular case of symmetry restoration. More specifically, we consider the restoration of angular momentum from an axially symmetric product state and extrapolate the use of the generalized Skyrme force to spin $I \neq 0$\footnote{This extrapolation concerns not only the fact that additional local densities should be considered for $I \neq 0$ as discussed in section~\ref{lda}, but also the fact that the underlying perturbative expansion has been written for the ground-state only~\cite{duguet}.}. The projected energy on spin $I$ and spin projection $M=0$ for an even-even nucleus is given by:

\begin{equation}
E^{\, n = 0}_{I0}  \, = \, \frac{\sum_{\alpha= -n}^{n} \, f^{I0}_{\alpha} \, \langle \, \Phi_{0}^{0} | \, \left[ t \, + \, v_{Skyrme}^{\left(0,\alpha\right)}\right] \, R (\alpha) \, | \, \Phi_{0}^{0}  \rangle}{\sum_{\alpha= -n}^{n} \,  f^{I0}_{\alpha} \, \langle \, \Phi_{0}^{0} | \, R (\alpha) \, | \,\Phi_{0}^{0}   \rangle} \, \, \, \, \, ,
\label{projavecskyrmemodif}
\end{equation}
where the coefficients of the mixing are:

\begin{equation}
f^{I0}_{\alpha} \,  = \, \frac{2I + 1}{2n} \, sin(\pi \alpha / n) \, d^{I \, \ast}_{00} (\pi \alpha / n) \, \, \, \, \, , \label{coeffprojJ}
\end{equation}
with $d^{I}_{M0}$ the Wigner function for the quantum numbers $(I,M,K=0)$~\cite{blaizot2}. The rotation operator for an angle $\pi \alpha / n$ around the $y$ axis orthogonal to the symmetry axis is:

\begin{equation}
R (\alpha) \, = \, e^{i \pi \alpha J_{y} / n} \, \, \, \, .
\label{rotop}
\end{equation}

In Eq.~\ref{projavecskyrmemodif},  $v_{Skyrme}^{\left(0,\alpha\right)}$ depends on the mean-field densities $\rho^{0}_{0} (\vec{R})$ and $\rho^{\alpha}_{0} (\vec{R})$ of the product state $| \, \Phi_{0}^{0}  \rangle$ and the rotated one $| \, \Phi^{\alpha}_{0}  \rangle \, = \, R (\alpha) \, | \, \Phi_{0}^{0}  \rangle$ as well as their mixed density $\rho^{\left(0,\alpha\right)}_{0} (\vec{R})$. For the derived prescription to be reliable for symmetry restoration, two properties have to be satisfied:

\bi
\item[$\bullet$] As for a bare Hamiltonian invariant under rotation, the projected energy~\ref{projavecskyrmemodif} has to be independent of the orientation axis in the laboratory frame\footnote{It is not necessary for $H^{\left(0,\alpha\right)}=t \, + \, v_{Skyrme}^{\left(0,\alpha\right)}$ to be invariant under rotation.}. 
\item[$\bullet$] The correlated energy~\ref{projavecskyrmemodif} has to be real\footnote{It is not necessary for $H^{\left(0,\alpha\right)}$ to be Hermitian.}.
\ei

In the present context, the first property is immediately satisfied. Indeed, the commutation of the projector $\hat{P}_{I0}$ with the bare Hamiltonian was done in the expression of the actual ground-state energy before any resummation and troncation took place in the perturbative expansion~\cite{duguet} and before the multi-body forces renormalization was performed. This is why the rotation matrix $R (\alpha)$ only appears on the r.h.s. of $v_{Skyrme}^{\left(0,\alpha\right)}$ in Eq.~\ref{projavecskyrmemodif}. Thus, $\langle \, \Phi_{0}^{0} | \, H^{\left(0,\alpha\right)} \, R (\alpha) \, | \, \Phi_{0}^{0}  \rangle$ only depends on the relative angle between the two involved product states. The projected energy is independent of the choice of the axis with respect to which angles are measured in the laboratory frame. The second property is demonstrated in appendix~\ref{verifprescription}. 

The prescription derived from the extended perturbation theory satisfies the minimal mathematical properties necessary to be used for symmetry restoration.

\section{Conclusions.}
\label{Conclu}

In this paper, we have analyzed the density-dependence of phenomenological effective two-body forces for calculations beyond the mean-field approximation. Up to now, two prescriptions have been used in the GCM and the projected mean-field method without being supported by strong theoretical arguments. They correspond to the local {\it mixed} density~\cite{bonche1,bonche2} and the local {\it correlated} density~\cite{anguiano1}. Similar results have been obtained with both in calculations dealing with projection on particle numbers~\cite{anguiano1}.

However, it is not clear whether other configuration mixing calculations such as projection on angular momentum for high spin states, projection on parity or GCM calculations for various collective coordinates involving extended nuclear shapes would give comparable results for both prescriptions. This remark is also relevant for physical properties involving exotic densities such as halos or neutron skins or for shape coexistence in nuclei. As a consequence, it appeared important to revisit this question. Instead of letting a quantitative agreement with experimental data decide which density should be used in this context, we have tried to come back to the origins of the density dependence of effective two-body forces.

First we have dealt with the Brueckner ladders summation which accounts for two-body correlations induced by the NN interaction in the presence of other nucleons. It is known to bring about a density dependence in the effective two-body force used together with a mean-field wave function. Developing an extended Brueckner-Goldstone scheme, we derived an effective force accounting for these correlations within the framework of mixed non-orthogonal Slater-determinants.

Discussing the foundations of a local approximation for the Brueckner matrix, we have explained why such a local expression has not been derived explicitly. As the imperfections and unnecessary complications of the formal derivation can be smoothed out through the use of phenomenological forces fitted to the data, we have taken the validity of such a local density approximation as an ansatz. We have deduced the corresponding functional of local mean-field densities by making it match the standard Skyrme or Gogny density-dependent term at the mean-field level. 

Second, we have analyzed the density dependence stemming from the possible renormalization of multi-body forces in the nuclear Hamiltonian. We have shown the formal equivalence between a three-body force and a two-body force depending linearly on the density for configuration mixing calculations, as soon as the mixed density is used. This result holds for the equations of motion obtained from the minimization of the energy, whatever the variational parameters are. The role of the rearrangement terms has been emphasized. Then, showing that a density dependence of the form $\rho^{\sigma}$ with a non integer value of $\sigma$, as used in Skyrme and Gogny phenomenological forces, can originate from the renormalization of multi-body force effects, we have generalized the result obtained for a three-body force.

Finally, two kinds of density dependence ought to be used depending on whether they deal with the renormalization of two-body correlations induced by the strong-repulsive core and the tensor part of the bare nucleon-nucleon interaction, or with the renormalization of multi-body force effects. One common feature of these two prescriptions is their dependence on the $N$-body matrix element in which they are inserted when expressing the approximate energy in terms of the mixed product states. They both coincide with the mean-field matter density when returning to the mean-field approximation. This shows how going beyond the mean field may open degrees of freedom in the effective force which are not fixed at the mean-field level. 

Using these results, we have proposed a theoretically grounded extension of the Skyrme force for configuration mixing calculations. Explaining in detail all the approximations performed on the way to this definition, we have discussed its possible generalizations. GCM and projected mean-field calculations testing presently proposed as well as existing prescriptions are now underway. Corresponding results are the aim of a forthcoming publication.

\section{Acknowledgments.}
\label{secremer}

One of the authors (T.D.) is indebted to U. Reinosa for fruitful discussions and to N. J. Hammond for the proof reading of the manuscript.

\begin{appendix}

\section{GCM and Dependence on $\rho^{\Psi}$.}
\label{calcgcmmixing}

In order to reproduce the effect of a three-body force, we incorporate the non-local correlated density:

\begin{equation}
\rho^{\Psi_{k}} (\vec{r}_{1}, \vec{r}_{2}) \, = \, \sum_{\alpha,\beta} \, f_{\beta}^{k \, \ast} f^{k}_{\alpha} \, \rho^{\left(\beta,\alpha\right)}_{0} (\vec{r}_{1}, \vec{r}_{2}) \, \, \, \, 
\label{corrcorr}
\end{equation}
into the two-body force defined by Eq.~\ref{defVeff2} and get the effective interaction:

\begin{eqnarray}
V^{(3)}_{eff} (\vec{r},\vec{R}) \, &=&  \, V^{(2)} (\vec{r}) \, + \, v (\vec{r}) \, \rho^{\Psi_{k}} (\vec{r}_{1}, \vec{r}_{2}) \,  \, \label{defVeff2corr} .
\end{eqnarray}

Using this density dependence, one gets a non-zero rearrangement term by varying the correlated energy $\langle \, \Psi_{k} | \, H^{(3)}_{eff}  | \, \Psi_{k}  \rangle / \langle \, \Psi_{k}  | \, \Psi_{k}  \rangle$ with respect to $f^{k \, \ast}_{\beta}$. This is different from the result obtained with the mixed density. The variation leads to the following equations of motion:

\begin{eqnarray}
&& \sum_{\alpha} \, f^{k}_{\alpha} \, \left[\sum_{i,j} \,  t_{ij} \, \rho^{\left(\beta,\alpha\right)}_{ji} \, \, +  \, \, \frac{1}{2} \,  \sum_{i,j,k,l} \, \overline{V}^{(2)}_{iljk} \,  \rho^{\left(\beta,\alpha\right)}_{ji} \, \rho^{\left(\beta,\alpha\right)}_{kl} \right.  \nb \\
 \label{equamotioncorr} \\
&& +    \frac{1}{6} \,  \sum_{i,j,k,l,m,n} \,  v_{ilnjkm} \, \rho^{\left(\beta,\alpha\right)}_{ji} \, \rho^{\left(\beta,\alpha\right)}_{kl} \,  \sum_{q} \, f^{k}_{q} \, \left( \rho^{\left(\beta,q\right)}_{mn} \, \frac{\langle \, \Phi^{\beta}_{0}  | \, \Phi^{q}_{0} \rangle}{\langle \, \Psi_{k}  | \, \Psi_{k} \rangle} \right.    \nb \\
&&  \nb \\
&& \hspace{5.5cm}+  \left. \left. \sum_{p} f^{\ast}_p \, \rho^{\left(p,q\right)}_{mn} \, \frac{\langle \, \Phi^{p}_{0}  | \, \Phi^{q}_{0} \rangle}{\langle \, \Psi_{k}  | \, \Psi_{k} \rangle} \right)   \right]   \langle \, \Phi^{\beta}_{0}  | \, \Phi^{\alpha}_{0} \rangle  \nb \\
&&\nb \\
&& \hspace{8cm} =  \, {\cal E}_{k}^{mix} \,  \sum_{\alpha} \, f^{k}_{\alpha} \,  \langle \, \Phi^{\beta}_{0}  | \, \Phi^{\alpha}_{0} \rangle \, \,  . \nb 
\end{eqnarray}

In order to reproduce Eq.~\ref{equamotion3} the term between parenthesis in Eq.~\ref{equamotioncorr} should be equal to $\rho^{\left(\beta,\alpha\right)}_{mn}$. This means that $\rho^{\Psi} + \partial \rho^{\Psi} / \partial f^{k \, \ast}_{\beta}$ should be equal to $\rho^{\left(\beta,\alpha\right)}$ for all $(\beta, \alpha)$. However, it be the case in general. Therefore, the use of the correlated density in the effective two-body force generates redundant terms. The above consideration suggests that it should be avoided in configuration mixing calculations.

\section{Projection and Three-Body Force.}
\label{calcprojmixing}

\subsection{Equation of Motion.}
\label{appequamotion}

The equations of motion defined by Eq.~\ref{equamotionproj} take the explicit form:

\begin{eqnarray}
&& \sum_{\beta,\alpha} \, f^{IM \, \ast}_{\beta} f^{IM}_{\alpha} \, \left[  \sum_{i,j} \,  t_{ij} \, \left( \rho^{\left(\beta,\alpha\right)}_{u'u} \rho^{\left(\beta,\alpha\right)}_{ji} -  \rho^{\left(\beta,\alpha\right)}_{u'i} \rho^{\left(\beta,\alpha\right)}_{ju} \right)  \right. \,  \nb \\
&& \nb \\
&& \hspace{0.7cm} +  \, \frac{1}{2} \,  \sum_{i,j,k,l} \, \overline{V}^{(2)}_{iljk} \,  \left( \rho^{\left(\beta,\alpha\right)}_{u'u} \rho^{\left(\beta,\alpha\right)}_{ji} \rho^{\left(\beta,\alpha\right)}_{kl} -  \rho^{\left(\beta,\alpha\right)}_{u'l} \rho^{\left(\beta,\alpha\right)}_{ji} \rho^{\left(\beta,\alpha\right)}_{ku} -  \rho^{\left(\beta,\alpha\right)}_{u'i} \rho^{\left(\beta,\alpha\right)}_{kl} \rho^{\left(\beta,\alpha\right)}_{ju}\right)   \nb \\
&& \label{expliciteqmotion} \\
&& \hspace{0.7cm}+ \,  \frac{1}{6} \,  \sum_{i,j,k,l,m,n} \, \overline{V}^{(3)}_{ilnjkm} \, \left( \rho^{\left(\beta,\alpha\right)}_{u'u} \rho^{\left(\beta,\alpha\right)}_{ji} \rho^{\left(\beta,\alpha\right)}_{kl} \rho^{\left(\beta,\alpha\right)}_{mn} -  \rho^{\left(\beta,\alpha\right)}_{u'l} \rho^{\left(\beta,\alpha\right)}_{ji} \rho^{\left(\beta,\alpha\right)}_{ku} \rho^{\left(\beta,\alpha\right)}_{mn} \right. \nb \\
&& \nb \\
&& \hspace{2.4cm}\left. \left. -  \rho^{\left(\beta,\alpha\right)}_{u'n} \rho^{\left(\beta,\alpha\right)}_{ji} \rho^{\left(\beta,\alpha\right)}_{kl}  \rho^{\left(\beta,\alpha\right)}_{mu} - \rho^{\left(\beta,\alpha\right)}_{u'i} \rho^{\left(\beta,\alpha\right)}_{ju} \rho^{\left(\beta,\alpha\right)}_{kl} \rho^{\left(\beta,\alpha\right)}_{mn}\right) \, \right]   \langle \, \Phi^{\beta}_{0} \, | \, \Phi^{\alpha}_{0} \rangle \nb \\
&& \nb \\
&& \hspace{6.7cm}= \, {\cal E}_{k}^{mix} \, \sum_{\beta,\alpha} \, f^{IM \, \ast}_{\beta} f^{IM}_{\alpha} \,\rho^{\left(\beta,\alpha\right)}_{u'u} \langle \, \, \Phi^{\beta}_{0} \, | \, \Phi^{\alpha}_{0} \rangle \, \, \, \, \, , \nb
\end{eqnarray}
for all $(u,u')$.

\subsection{Rearrangement Term.}
\label{reaterm}

For the density-dependent Hamiltonian $H^{(3)(\beta,\alpha)}_{eff}$, Eq.~\ref{equamotionproj} must be modified in order to include the rearrangement term originating from the variation of the two-body interaction with respect to the individual wave-functions. The equations of motion become:

\begin{eqnarray}
\sum_{\beta,\alpha} \, f^{IM \, \ast}_{\beta} f^{IM}_{\alpha} \left[\langle \, \Phi^{\beta}_{0}  | \, c^{\dagger}_{u} \, c_{u'} \,  H^{(3)(\beta,\alpha)}_{eff} \, | \,\Phi^{\alpha}_{0}   \rangle  + 2 \, \langle \, \Phi^{\beta}_{0}  |  \, \frac{\partial V^{(3)(\beta,\alpha)}_{eff} (\vec{r},\vec{R})}{\partial \rho^{\left(\beta,\alpha\right)} (\vec{r}_{1}, \vec{r}_{2})} \, \,  \frac{\partial \rho^{\left(\beta,\alpha\right)} (\vec{r}_{1}, \vec{r}_{2})}{\partial  Z^{I}_{uu'}} \, | \,\Phi^{\alpha}_{0}  \rangle\right] \, && \nb \\
&& \nb \\
\!= \, {\cal E}_{k}^{mix} \, \langle \, \Psi_{k} \, | \, c^{\dagger}_{u} \, c_{u'} \, | \, \Psi_{k} \rangle \, && \, \, \, \, \, \, \, \, . \label{eqmotionreaterm}
\end{eqnarray}

The calculation of the rearrangement term requires the evaluation of:

\begin{eqnarray}
\frac{\partial \rho^{\left(\beta,\alpha\right)} (\vec{r}_{1}, \vec{r}_{2})}{\partial  Z^{I}_{uu'}} &=& \frac{1}{2} \, \left[ \frac{\langle \, \Phi^{\beta}_{0}  |  \, c^{\dagger}_{u} \, c_{u'} \,  \hat{\rho} (\vec{r}_{1}, \vec{r}_{2}) \,| \,\Phi^{\alpha}_{0}   \rangle }{\langle \, \Phi^{\beta}_{0} | \,\Phi^{\alpha}_{0}   \rangle} \, - \, \frac{\langle \, \Phi^{\beta}_{0}  |  \, c^{\dagger}_{u} \, c_{u'} \,| \,\Phi^{\alpha}_{0}  \rangle \, \langle \, \Phi^{\beta}_{0}  | \,  \hat{\rho} (\vec{r}_{1}, \vec{r}_{2}) \,| \,\Phi^{\alpha}_{0}  \rangle}{\langle \, \Phi^{\beta}_{0} | \,\Phi^{\alpha}_{0}  \rangle^2} \right] \nb \\
&& \nb \\
&=& - \frac{1}{2} \, \sum_{i,j} \, \varphi_{I}^{\ast} (\vec{r}_{2}, \zeta_{z}', s_{z}', t') \, \varphi_{J} (\vec{r}_{1}, \zeta_{z}, s_{z}, t) \, \rho^{\left(\beta,\alpha\right)}_{u'i} \, \rho^{\left(\beta,\alpha\right)}_{ju}  \, \, \, \, \, . \label{diffmixeddens}
\end{eqnarray}

We sum the contribution from the density-dependent part of $V^{(3)(\beta,\alpha)}_{eff}$ in the first term of Eq.~\ref{eqmotionreaterm} together with the rearrangement term, and obtain the total contribution:

\begin{eqnarray}
&& \frac{1}{4} \, \sum_{\beta,\alpha} \, f^{IM \, \ast}_{\beta} f^{IM}_{\alpha} \, \sum_{i,j,k,l} \, \overline{\langle \varphi_{i} \, \varphi_{l} \, |  \, v (\vec{r}_{1}-\vec{r}_{2}) \, \rho^{\left(\beta,\alpha\right)} (\vec{r}_{1}, \vec{r}_{2}) \,  | \, \varphi_{k} \, \varphi_{j}  \rangle} \, \langle \, \Phi^{\beta}_{0} \, | \, c^{\dagger}_{u} \, c_{u'} \, c^{\dagger}_{i} \, c^{\dagger}_{l} \, c_{k} \, c_{j}\, | \, \Phi^{\alpha}_{0}  \rangle  \,  \nb \\ 
&& \nb \\ 
&&+ \, \frac{1}{2} \, \sum_{\beta,\alpha} \, f^{IM \, \ast}_{\beta} f^{IM}_{\alpha} \, \sum_{i,j,k,l} \, \overline{\langle \varphi_{i} \, \varphi_{l} \, | \, v (\vec{r}_{1}-\vec{r}_{2}) \, \frac{\partial \rho^{\left(\beta,\alpha\right)} (\vec{r}_{1}, \vec{r}_{2})}{\partial  Z^{I}_{uu'}} \, | \, \varphi_{k} \, \varphi_{j}  \rangle} \,  \langle \, \Phi^{\beta}_{0} \, | \, c^{\dagger}_{i} \, c^{\dagger}_{l} \, c_{k} \, c_{j}\, | \, \Phi^{\alpha}_{0}  \rangle  \,  \nb \\ 
&& \nb \\ 
&&= \, \frac{1}{6} \, \sum_{\beta,\alpha} \, f^{IM \, \ast}_{\beta} f^{IM}_{\alpha}  \,  \sum_{i,j,k,l,n,m} \, v_{ilnjkm}  \left( \rho^{\left(\beta,\alpha\right)}_{u'u} \rho^{\left(\beta,\alpha\right)}_{ji} \rho^{\left(\beta,\alpha\right)}_{kl} \rho^{\left(\beta,\alpha\right)}_{mn} -  \rho^{\left(\beta,\alpha\right)}_{u'l} \rho^{\left(\beta,\alpha\right)}_{ji} \rho^{\left(\beta,\alpha\right)}_{ku} \rho^{\left(\beta,\alpha\right)}_{mn} \right. \nb \\
&& \label{sum} \\
&& \hspace{7cm} \left. - \rho^{\left(\beta,\alpha\right)}_{u'i} \rho^{\left(\beta,\alpha\right)}_{ju} \rho^{\left(\beta,\alpha\right)}_{kl} \rho^{\left(\beta,\alpha\right)}_{mn}\right) \langle \, \Phi^{\beta}_{0} \, | \, \Phi^{\alpha}_{0} \rangle \nb \\
&& \nb \\
&& - \frac{1}{6} \, \sum_{\beta,\alpha} \, f^{IM \, \ast}_{\beta} f^{IM}_{\alpha}   \,  \sum_{i,j,k,l,n,m} \, v_{ilnjkm} \, \rho^{\left(\beta,\alpha\right)}_{u'n} \rho^{\left(\beta,\alpha\right)}_{ji} \rho^{\left(\beta,\alpha\right)}_{kl}  \rho^{\left(\beta,\alpha\right)}_{mu} \langle \, \Phi^{\beta}_{0} \, | \, \Phi^{\alpha}_{0} \rangle \nb \\
&& \nb \\
&& = \frac{1}{6}  \, \sum_{\beta,\alpha} \, f^{IM \, \ast}_{\beta} f^{IM}_{\alpha}   \,  \sum_{i,j,k,l,n,m} \, v_{ilnjkm}  \left( \rho^{\left(\beta,\alpha\right)}_{u'u} \rho^{\left(\beta,\alpha\right)}_{ji} \rho^{\left(\beta,\alpha\right)}_{kl} \rho^{\left(\beta,\alpha\right)}_{mn} -  \rho^{\left(\beta,\alpha\right)}_{u'l} \rho^{\left(\beta,\alpha\right)}_{ji} \rho^{\left(\beta,\alpha\right)}_{ku} \rho^{\left(\beta,\alpha\right)}_{mn}  \right. \nb \\
&& \nb \\
&& \hspace{3cm} \left. - \rho^{\left(\beta,\alpha\right)}_{u'n} \rho^{\left(\beta,\alpha\right)}_{ji} \rho^{\left(\beta,\alpha\right)}_{kl}  \rho^{\left(\beta,\alpha\right)}_{mu} - \rho^{\left(\beta,\alpha\right)}_{u'i} \rho^{\left(\beta,\alpha\right)}_{ju} \rho^{\left(\beta,\alpha\right)}_{kl} \rho^{\left(\beta,\alpha\right)}_{mn}\right) \langle \, \Phi^{\beta}_{0} \, | \, \Phi^{\alpha}_{0} \rangle \, \, \, . \nb
\end{eqnarray}
where $v$ takes the form $v (\vec{r}_{1}-\vec{r}_{2}, \vec{p}, \hat{\vec{\sigma}}_{1}, \hat{\vec{\sigma}}_{2}, \hat{\vec{\tau}}_{1}.\hat{\vec{\tau}}_{2})$ and where its matrix elements $v_{ilnjkm}$ are defined through Eq.~\ref{matrixelementeff}.

The comparison with Eq.~\ref{expliciteqmotion} shows that the same formal contributions to the equations of motion as the one coming from a three-body force  are obtained as soon as $v_{ilnjkm}$ is able to reproduce $\overline{V}^{(3)}_{ilnjkm}$. The rearrangement term is essential as it gives a term with a combination of indices which cannot be obtained through the redefinition of the two-body force only.

\section{Projection and Dependence on $\rho^{\Psi}$.}
\label{argument}

In Ref.~\cite{val2}, Valor and collaborators gave an argument in favor of the correlated density. They argued that once the correlated energy is developed in terms of product functions as given by Eq.~\ref{energy}, the introduction of a dependence of the effective Hamiltonian $H_{eff}$ on the mixing angles $(\alpha, \beta)$ as it is the case when using the mixed density for instance prevents from extracting the mean energy of the correlated state with good quantum numbers. It seems to favor of $\rho^{\Psi_{k}}$ which is independent of the mixing angles.

Let us exemplify the situation through the projection of an HFB wave-function on good particle number~\cite{ring1}:

\begin{eqnarray}
| \, \Psi_{N}  \rangle \, &=& \, \hat{P}_{N} \, | \, \Phi^{0}_{0}  \rangle \, \, \, ,\nb \\
&& \label{projproj} \\
\hat{P}_{N} \, &=& \, \frac{1}{2\pi} \int_{-\pi}^{\pi} \, d\alpha \, e^{i \alpha (\hat{N} - N)} \nb \, \, \, ,
\end{eqnarray}
where $\hat{N}$, $N$ and $\alpha$ are respectively the particle number operator, the actual number of particles and the mixing angle in gauge space. For simplicity, a single kind of nucleons is considered here.

First, the energy of the correlated state $| \, \Psi_{N}  \rangle$ is developed in terms of the mixed product states $| \, \Phi^{\alpha}_{0}  \rangle \, = \, e^{i \alpha \hat{N}} \, | \, \Phi^{0}_{0}  \rangle$:

\begin{equation}
{\cal E}^{mix}_{N} \, = \, \frac{\int_{-\pi}^{\pi} \, d\alpha \, e^{-i \alpha N} \, \langle \, \Phi^{0}_{0} \, | \, H_{eff} \,  e^{i \alpha \hat{N}} \,  | \, \Phi^{0}_{0}  \rangle}{\int_{-\pi}^{\pi} \, d\alpha \, e^{-i \alpha N} \,  \langle \, \Phi^{0}_{0} \, | \,e^{i \alpha \hat{N}} \,  | \, \Phi^{0}_{0}  \rangle} \, \, \, \, \, \, ,
\label{effectiveness}
\end{equation}
which clearly shows how the projection picks up the energy associated with the component of $| \, \Phi^{0}_{0}  \rangle$ having exactly $N$ particles. Then, if one makes $H_{eff}$ depend on $\alpha$, the calculated energy will not be expressible as a mean value $\langle \, \Psi_{N} \, | \, H_{eff} \, | \, \Psi_{N}  \rangle / \langle \, \Psi_{N} \, | \, \Psi_{N}  \rangle$ in the projected state having the good quantum number $N$. This is correct but not pertinent here. In order to understand why, one has to go back to the origin of $H_{eff}$'s effectiveness characterized by its density dependence. 

First, one has to be aware that the rational of any microscopic calculation (variational or perturbative) is always to approximate the actual eigenstates and eigenenergies of the system, the leading quantity being the energy. Thus, coming back to our example, the ultimate goal is not to obtain the mean-value of some effective Hamiltonian in the projected state but to reproduce as closely as possible the eigenenergy:

\begin{equation}
\frac{\langle \, \Theta_{0} \, | \, H \, | \, \Theta _{0} \rangle}{\langle \, \Theta_{0} \, | \, \Theta_{0}  \rangle} \, \, \, \, \, \, ,
\label{actual}
\end{equation}
where $H$ is the actual Hamiltonian of the system and  $| \, \Theta _{0} \rangle$ the unknown ground-state wave-function. 

Within the projected mean-field method, this is done through an approximation as given by Eq.~\ref{effectiveness} where $H_{eff}$ is effective in order to remove the repulsive core of the bare nucleon-nucleon interaction and/or to renormalize multi-body forces effect. It has been shown in this context how $H_{eff}$ should depend on $\alpha$ and why the corresponding energy ${\cal E}^{mix}_{N}$ could not be factorized into $\langle \, \Psi_{N} \, | \, H_{eff} \, | \, \Psi_{N}  \rangle / \langle \, \Psi_{N} \, | \, \Psi_{N}  \rangle$.

For instance, considering only the renormalization of multi-body forces effects, the argument given in Ref.~\cite{val2} omits that if one wants to reproduce the projected energy $\langle \, \Psi_{0} \, | \, H^{(3)} \, | \, \Psi_{0}  \rangle / \langle \, \Psi_{0} \, | \, \Psi_{0}  \rangle$ including multi-body forces, itself approximating $\langle \, \Theta_{0} \, | \, H^{(3)} \, | \, \Theta_{0}  \rangle / \langle \, \Theta_{0} \, | \, \Theta_{0}  \rangle$. Using an effective two-body Hamiltonian $H^{(3)}_{eff}$ to do so, it is necessary to make this latter depend on $\alpha$. Then, the impossibility to factorize the energy~\ref{effectiveness} does not contradict the fact that $| \, \Psi_{0}  \rangle$ remains the corresponding approximate state of the system from which other observables can be evaluated.

\section{$v_{Skyrme}^{\left(\beta,\alpha\right)}$ for Symmetry Restoration.}
\label{verifprescription}

We have to check whether the projected ground-state energy~\ref{projavecskyrmemodif} is real. Thanks to the symmetric integration around 0 on the variable $\alpha$ in Eq.~\ref{projavecskyrmemodif}, it is sufficient to prove that:

\begin{equation}
\left[\langle \, \Phi_{0}^{0} | \, H^{\left(0,\alpha\right)} \, R (\alpha) \, | \, \Phi_{0}^{0}  \rangle\right]^{\ast} \, = \, \langle \, \Phi_{0}^{0} | \, H^{\left(0,-\alpha\right)} \, R (-\alpha) \, | \, \Phi_{0}^{0}  \rangle  \, \, \, \, \, \, \, ,
\label{rot0}
\end{equation}
is valid for the mixed Hamiltonian $H^{\left(0,\alpha\right)}$ specified to the projection on angular momentum. The same property is straightforward for the overlap $\langle \, \Phi_{0}^{0} | \, R (\alpha) \, | \, \Phi_{0}^{0}  \rangle$.

Let us first look at the different densities under rotation. We introduce the unitary $3*3$ matrix ${\cal R} (\alpha)$ which rotates an eigenvector of the position operator:

\begin{equation}
R (\alpha) \, | \, \vec{r} \, \rangle \, \equiv \, | \, {\cal R} (\alpha) \, \vec{r} \, \rangle  \, \, \, \, \, \, \, \, ,
\label{rot1}
\end{equation}
or a vector operator, such as the vector position operator:

\begin{equation}
R^{\dagger} (\alpha) \, \vec{r} \, R (\alpha) \, = \, {\cal R} (\alpha) \, \vec{r}  \, \, \, \, \, \, \, \, .
\label{rot2}
\end{equation}

The local scalar-isoscalar part of the mixed density specified to the projection is:

\begin{equation}
\rho^{\left(\beta,\alpha\right)}_{0} (\vec{r}) \, \, = \, \, \frac{\langle \, \Phi^{0}_{0} \, | \, R^{\dagger} (\beta) \, \hat{\rho}_{0} (\vec{r}) \, R (\alpha) \,  | \, \Phi^{0}_{0}  \rangle}{\langle \, \Phi^{0}_{0} \, | \, R^{\dagger} (\beta) \, R (\alpha) \,  | \, \Phi^{0}_{0}  \rangle} \, \, \, \, \, \, \, \, .
\label{mixed2}
\end{equation}

This quantity is a matrix element between two $N$-body states where $\vec{r} \,$ is an external variable. As an operator function of the vector position operator $\vec{r}$, its behavior under rotation is, thanks to Eq.~\ref{rot2}:

\begin{equation}
R^{\dagger} (\alpha) \, \rho^{\left(\beta,\alpha\right)}_{0} (\vec{r}) \, R (\alpha) \, = \, \rho^{\left(\beta,\alpha\right)}_{0} ({\cal R} (\alpha) \, \vec{r})  \, \, \, \, \, \, \, \, .
\label{rot3}
\end{equation}

As an operator function of the positions $\vec{r}_{i}$ of the nucleons, the transformation under rotation of the local scalar-isoscalar part of the one-body density operator can be written as:

\begin{equation}
R^{\dagger} (\alpha) \, \hat{\rho}_{0} (\vec{r}) \, R (\alpha) \, = \, \sum_{i=1}^{N} \, \delta (\vec{r} - {\cal R} (\alpha) \, \vec{r}_{i}) \, = \, \sum_{i=1}^{N} \, \delta ({\cal R}^{\dagger} (\alpha) \, \vec{r} - \vec{r}_{i}) \, = \, \hat{\rho}_{0} ({\cal R}^{\dagger} (\alpha) \, \vec{r}) \, \, \, \, .
\label{rot4}
\end{equation}

Following Eq.~\ref{rot3}, one can write:

\begin{equation}
R^{\dagger} (\gamma) \, \rho^{\left(\beta,\alpha\right)}_{0} (\vec{r}) \, = \, R^{\dagger} (\gamma) \, \rho^{\left(\beta,\alpha\right)}_{0} (\vec{r}) \, R (\gamma) \, R^{\dagger} (\gamma) \, = \, \rho^{\left(\beta,\alpha\right)}_{0} ({\cal R} (\gamma) \, \vec{r}) \, R (-\gamma) \, ,
\label{rot5}
\end{equation}
which, thanks to Eq.~\ref{mixed2} and~\ref{rot4}, can be recast as~:

\begin{eqnarray}
R^{\dagger} (\gamma) \, \rho^{\left(\beta,\alpha\right)}_{0} (\vec{r}) \, &=& \, \frac{\langle \, \Phi^{0}_{0} \, | \, R^{\dagger} (\beta) \, \hat{\rho}_{0} ({\cal R} (\gamma) \, \vec{r}) \, R (\alpha) \,  | \, \Phi^{0}_{0}  \rangle}{\langle \, \Phi^{0}_{0} \, | \, R^{\dagger} (\beta) \, R (\alpha) \,  | \, \Phi^{0}_{0}  \rangle} \, \, R (-\gamma) \nb \\
&& \nb \\
&=& \, \frac{\langle \, \Phi^{0}_{0} \, | \, R^{\dagger} (\beta) \, R (\gamma) \, \hat{\rho}_{0} (\vec{r}) \, R^{\dagger} (\gamma) \, R (\alpha) \,  | \, \Phi^{0}_{0}  \rangle}{\langle \, \Phi^{0}_{0} \, | \, R^{\dagger} (\beta) \, R (\gamma) \, R^{\dagger} (\gamma) \, R (\alpha) \,  | \, \Phi^{0}_{0}  \rangle} \, \, R (-\gamma)  \label{rot6} \\
&& \nb \\
&=& \, \rho^{\left(\beta - \gamma, \alpha - \gamma\right)}_{0} (\vec{r}) \, R (-\gamma) \, \, \, . \nb
\end{eqnarray}

We also need to calculate the complex conjugate of the local scalar-isoscalar part of the mixed density:

\begin{eqnarray}
\left(\rho^{\left(\beta,\alpha\right)}_{0} (\vec{r})\right)^{\ast} \, &=& \, \frac{\langle \, \Phi^{0}_{0} \, | \, R^{\dagger} (\alpha) \, \hat{\rho}_{0} (\vec{r}) \, R (\beta) \,  | \, \Phi^{0}_{0}  \rangle}{\langle \, \Phi^{0}_{0} \, | \, R^{\dagger} (\alpha) \, R (\beta) \,  | \, \Phi^{0}_{0}  \rangle} \, \nb \\
&& \label{rot9} \\
&=& \, \rho^{\left(\alpha,\beta\right)}_{0} (\vec{r}) \, \, \, , \nb
\end{eqnarray}
as well as the structure of the effective Hamiltonian deduced from Eq.~\ref{prescription2}:

\begin{equation}
H^{\left(0,\alpha\right)} \, = \, \frac{X}{2} \, \left(H \left[\rho_{0}^{0} (\vec{R})\right] \, + \,  H \left[\rho_{0}^{\alpha} (\vec{R})\right]\right) \, + \, (1-X) \, H \left[\rho^{\left(0,\alpha\right)}_{0} (\vec{R})\right]  \, \, \, ,
\label{ham}
\end{equation}
where $H$ is the mean-field (Skyrme) Hamiltonian. Finally, we can now prove the identity~\ref{rot0}:
 
\begin{eqnarray}
&& \left[\langle \, \Phi_{0}^{0} | H^{\left(0,\alpha\right)} R (\alpha) | \, \Phi_{0}^{0}  \rangle\right]^{\ast}  \nb \\
&& \nb \\
&=& \langle \, \Phi_{0}^{0} \, | R^{\dagger} (\alpha) \left\{\frac{X}{2}\! \left(H \left[(\rho_{0}^{0} (\vec{R}))^{\ast}\right]\! +\! H \left[(\rho_{0}^{\alpha} (\vec{R}))^{\ast}\right]\right)\! +\! (1\!-\!X) H \left[(\rho^{\left(0,\alpha\right)}_{0} (\vec{R}))^{\ast}\right]\right\} | \, \Phi_{0}^{0}  \rangle \nb \\
&& \nb \\
&=& \langle \, \Phi_{0}^{0} \, | R^{\dagger} (\alpha) \left\{\frac{X}{2}\! \left(H \left[\rho_{0}^{0} (\vec{R})\right]\! +\! H \left[\rho_{0}^{\alpha} (\vec{R})\right]\right)\! +\! (1\!-\!X) H \left[\rho^{\left(\alpha,0\right)}_{0} (\vec{R})\right]\right\} | \, \Phi_{0}^{0}  \rangle  \nb \\
&& \label{rot7} \\
&=& \langle \, \Phi_{0}^{0} \, | \left\{\frac{X}{2}\! \left(H \left[\rho_{0}^{-\alpha} (\vec{R})\right]\! +\! H \left[\rho_{0}^{0} (\vec{R})\right]\right)\! +\! (1\!-\!X) H \left[\rho^{\left(0,-\alpha\right)}_{0} (\vec{R})\right]\right\} R (-\alpha) | \, \Phi_{0}^{0}  \rangle \nb \\
&& \nb \\
&=& \langle \, \Phi_{0}^{0} | H^{\left(0,-\alpha\right)} R (-\alpha) | \, \Phi_{0}^{0}  \rangle \, \, , \nb
\end{eqnarray}
where we have applied Eq.~\ref{rot9} and~\ref{rot6} to the three densities involved. The extension to triaxially deformed product states poses no difficulty. 

\end{appendix}

\clearpage


\end{document}